\documentclass[journal,onecolumn,draftclsnofoot,11pt]{IEEEtran}
\usepackage{acronym}
\usepackage{amsfonts}
\usepackage{times}
\usepackage{graphicx}
\usepackage{latexsym}
\usepackage{dsfont}
\usepackage{amssymb}
\usepackage{amsmath}
\usepackage{cite}
\usepackage{verbatim}
\usepackage{subfigure}

\newcommand{\Ns}{N_{\text{s}}}

\newcommand{\bHb}{\mathbf{H}_{\text{b}}}

\newcommand{\fig}[1]{Fig.\ \ref{#1}}

\newcommand{\tr}{\text{tr}}

\def\bb0{{\mathbb{0}}}

\def\ba{{\mathbf{a}}}
\def\bb{{\mathbf{b}}}

\def\bff{{\mathbf{f}}}

\def\bh{{\mathbf{h}}}

\def\bm{{\mathbf{m}}}
\def\bn{{\mathbf{n}}}

\def\bs{{\mathbf{s}}}

\def\bv{{\mathbf{v}}}
\def\bw{{\mathbf{w}}}
\def\bx{{\mathbf{x}}}
\def\by{{\mathbf{y}}}

\def\b0{{\mathbf{0}}}

\def\bA{{\mathbf{A}}}
\def\bB{{\mathbf{B}}}

\def\bF{{\mathbf{F}}}
\def\bG{{\mathbf{G}}}
\def\bH{{\mathbf{H}}}
\def\bI{{\mathbf{I}}}

\def\bQ{{\mathbf{Q}}}
\def\bR{{\mathbf{R}}}

\def\bU{{\mathbf{U}}}
\def\bV{{\mathbf{V}}}
\def\bW{{\mathbf{W}}}
\def\bX{{\mathbf{X}}}
\def\bY{{\mathbf{Y}}}
\def\bZ{{\mathbf{Z}}}

\def\bbC{{\mathbb{C}}}

\def\bbE{{\mathbb{E}}}

\def\cA{\mathcal{A}}

\def\cF{\mathcal{F}}

\def\cM{\mathcal{M}}
\def\cN{\mathcal{N}}
\def\cO{\mathcal{O}}

\def\sf0{{\mathsf{0}}}

\def\Nt{{N_\text{t}}}
\def\Nr{{N_\text{r}}}
\def\Ns{{N_\text{s}}}

\def\aT{{\ba_{\text{T}}}}
\def\aR{{\ba_{\text{R}}}}
\def\Np{{N_\text{p}}}

\newcommand{\Mt}{M_{\text{t}}}
\newcommand{\Mr}{M_{\text{r}}}

\newcommand{\secref}[1]{{Section}~\ref{#1}}
\usepackage{epstopdf}

\usepackage{color}

\definecolor{purple(x11)}{rgb}{0.63, 0.36, 0.94}
\definecolor{cadmiumgreen}{rgb}{0.0, 0.42, 0.24}

\newcommand{\Lt}{L_{\text{t}}}
\newcommand{\Lr}{L_{\text{r}}}
\newcommand{\peff}{p_{\text{eff}}}

\DeclareMathOperator*{\argmin}{arg\,min}

\begin{document}

\title{An Overview of Signal Processing Techniques for Millimeter Wave MIMO Systems}
\author{Robert W. Heath Jr., Nuria Gonzalez-Prelcic, Sundeep Rangan, Wonil Roh, and Akbar Sayeed
\thanks{R. W. Heath Jr. is with The University of Texas at Austin, Austin, TX, USA (email: rheath@utexas.edu). 
Nuria Gonzalez-Prelcic is with the University of Vigo, Spain, (email: nuria@gts.uvigo.es). 
Sundeep Rangan is with New York University, USA, (email: srangan@nyu.edu).
Wonil Roh, is with Samsung Electronics, South Korea, (email: wonil.roh@samsung.com).
Akbar Sayeed, is with the University of Wisconsin-Madison, USA, (email: akbar@engr.wisc.edu). R. Heath would like to acknowledge support from the National Science Foundation under grant numbers NSF-CCF-1319556, NSF-CCF-1514275, and NSF-CCF-1527079, the U.S. Department of Transportation through the Data-Supported Transportation Operations and Planning (D-STOP) Tier 1 University Transportation Center, the Intel / Verizon 5G program, MERL, Nokia, Huawei, and Toyota. A. Sayeed would like to acknowledge support from the National Science Foundation under grant numbers ECCS-1247583 and IIP-1444962, and the Wisconsin Alumni Research Foundation.
}}

\maketitle

\begin{abstract}
Communication at millimeter wave (mmWave) frequencies is defining a new era of wireless communication. The mmWave
band offers higher bandwidth communication channels versus those presently used in commercial wireless systems. The applications of mmWave are immense: wireless local and personal area networks in the unlicensed band, 5G cellular systems, not to mention vehicular area networks, ad hoc networks, and wearables. Signal processing is critical for enabling the next generation of mmWave communication. Due to the use of large antenna arrays at the transmitter and receiver, combined with radio frequency and mixed signal power constraints, new  multiple-input multiple-output (MIMO) communication signal processing techniques are needed. Because of the wide bandwidths, low complexity transceiver algorithms become important. There are opportunities to exploit techniques like compressed sensing for channel estimation and beamforming. This article provides an overview of signal processing challenges in mmWave wireless systems, with an emphasis on those faced by using MIMO communication at  higher carrier frequencies. 
\end{abstract}

\section{Introduction} \label{sec:intro}

\IEEEPARstart{T}{he}  millimeter wave (mmWave) band is the frontier for commercial -- high volume consumer -- wireless communication systems \cite{RapHeaBook}. MmWave makes use of spectrum from $30$ GHz to $300$ GHz whereas most consumer wireless systems operate at carrier frequencies below $6$ GHz. The main benefit of going to mmWave carrier frequencies is the larger spectral channels. For example, channels with $2$ GHz of bandwidth are common for systems operating in the $60$ GHz unlicensed mmWave band. Larger bandwidth channels mean higher data rates. Despite the recent interest in mmWave, the study of mmWave is in fact as old as wireless itself. Some of the first experiments like those of Bose  and Lebedev \cite{Emerson1997} were performed in the 1890s in  the mmWave band.


The first standardized consumer radios were in the $60$ GHz unlicensed band. WirelessHD \cite{WirelessHD} is the name for the successful personal area network (PAN) technology developed by a consortium of companies. It is used primarily to replace cables that carry uncompressed high definition video. IEEE 802.11ad \cite{80211ad} is a wireless local area network (WLAN) standard. It was essentially developed in the former WiGig consortium that was later absorbed into the WiFi Alliance. The development of wireless communication in the $60$ GHz unlicensed band was the topic of tremendous amounts of research \cite{Giannetti:99,ted2,Bourdoux2006,Smulders2007,Dan07,Sing09,Dan10,Torkildson,Singh11}. The aforementioned PAN and LAN standards use about 2 GHz of bandwidth and support OFDM (orthogonal frequency division multiplexing) or SC-FDE (single-carrier frequency-domain equalization)  type modulations to provide data rates up to 6 Gbps. Beamforming through several (up to four) small antenna arrays is also supported. Evolutions of these standards are expected to support more sophisticated forms of multiple-input multiple-output (MIMO) communication for higher data rates. Products based on WirelessHD have been available for several years while those based on IEEE 802.11ad are starting to ship in higher volumes. 
It seems that WLAN and PAN devices operating at $60$ GHz will be the first widely deployed consumer wireless devices at mmWave.  

MmWave is also receiving tremendous interest by academia, industry, and government for 5G cellular systems \cite{Pi_Zhouyue_MCOM11,Rappaport_Access13,Akdeniz:14,Roh2014,BaiHeath:CoverageRate:2015,FCC}. The main reason is that spectrum available in sub-6 GHz bands is limited. Though signal processing approaches like cognitive radio \cite{Mitola99,Haykin2005} free more spectrum, it still is not enough if gigabit-per-second data rates are required. Initial work has established the viability of 5G cellular through propagation studies and later through system capacity analysis. Surprisingly, there is much earlier work on mmWave cellular which proposes the integration of voice/data communication at 60 GHz \cite{Walke85}. The Federal Communication Commission in the USA is among the first to back enthusiasm behind 5G with spectrum for mobile cellular applications \cite{FCC}.

MmWave is already a significant footprint wireless backhaul. Traditional physical layer designs for 60 GHz backhaul assume expensive directional antennas, reducing cost advantages over wired solutions \cite{RapHeaBook}. Low cost mmWave technologies with adaptive arrays, however,  are actively being developed to backhaul densely distributed small cells  in urban environments. In this scenario, distances are very short but the operating expenditures associated with using fiber optical cable may still be prohibitive. It will be possible to establish high capacity connections using state-of-art, low cost mmWave devices \cite{Hur2013,Dehos2014}. Self-backhaul may even be possible in millimeter wave cellular systems \cite{SinghEtAl15}.


MmWave has other potential applications as well. For example, with the recent excitement related to connected and autonomous vehicles, mmWave may play a role in providing high data rate connections between cars. This is natural because mmWave is already the backbone of automotive radar, which has been widely deployed and developed over the past ten years \cite{Meinel2013}. The combination of mmWave communication and radar \cite{Sturm2011} is also interesting for mmWave applications. MmWave could be used to enable high rate low latency connections to clouds that permit remote driving of vehicles through new mmWave vehicle-to-infrastructure applications. MmWave is also of interest for high speed wearable networks that connect cell phone, smart watch, augmented reality glasses, and virtual reality headsets \cite{Pyattaev2015}. Clearly the future is bright for new applications of mmWave.

Signal processing is of critical importance for millimeter wave cellular systems. The reasons why  signal processing is different in millimeter wave frequencies than at lower frequencies \cite{AMGH2014,brady:taps12} are: (i) there are new constraints on the hardware in part due to the high frequency and bandwidth communication channels,  (ii) the channel models are different, and (iii) large arrays will be used at both the transmitter and receivers. These differences underly the foundations of this survey article.

New hardware constraints arise from practical considerations like power consumption and circuit technology. One signal processing implication is renewed interest in 
 partitioning signal processing operations between analog and digital domains to reduce, for example, the number of analog-to-digital converters or their resolution. This has led to the development of hybrid beamforming architectures \cite{sayeed:all10,brady:taps12,ElAyach_2014,Alkhateeb_JSTSP14,Han2015}, beamspace signal processing techniques \cite{Bajwa2010,sayeed:02a}, lens-based analog beamforming antennas \cite{brady:taps12}, and low-rate ADC methods \cite{Mo_Jianhua_ITA14,Mo_Jianhua_Asilomar14}. Another signal processing implication is that analog components like phase shifters are imperfect (quantized phase and insertion loss). This leads to new mathematical models of impairments, new analyses of the effects of these impairments, and new algorithms that yield good performance even in the presence of impairments. We identify several of the signal processing challenges that arise from hardware constraints in this article.


The channel models at mmWave are different because the propagation environment has a different effect on smaller wavelength signals \cite{RapHeaBook}. For example, diffraction tends to be lower due to the reduced Fresnel zone, scattering is higher due to the increased effective roughness of materials, and penetration losses can be much larger. Mmwave channel models use some common properties as low frequency systems (multi-path delay spread, angle spread, and Doppler shift), with different parameters though (few and clustered paths for example leading to more sparsity in the channel). In addtion, some new features are introduced as well to account for  high sensitivity to blockages (buildings, human body, or fingers) and  strong differences between line-of-sight and non-line-of-sight propagation conditions. There are many opportunities to exploit the mathematical properties of sparsity in channel estimation and equalization and precoder/combiner design. 


The arrays discussed for mmWave communication may be  large. Example array sizes in the literature include  16 elements in \cite{Cudak2014} or 256 elements in \cite{Rebeiz2015}, but the arrays may even be larger at the base station in a cellular system. IEEE 802.11ad products with 32 elements are already commercially available. To provide sufficient link margin, in most mmWave communication systems, arrays will be used at both the transmitter and receiver, creating many opportunities to apply MIMO communication techniques. The MIMO techniques applied will be different though due to the different channel characteristics and additional hardware constraints found at mmWave frequencies. The connection between MIMO and mmWave is the main reason that we emphasize signal processing for mmWave MIMO systems. 

The combined implications of hardware constraints, channel models, and large arrays has a far-reaching impact on the design of mmWave communication systems. For example, mmWave cellular systems might have new architectural features. For example, devices might maintain active connections with multiple base stations to achieve diversity from building, human, or self-body blockages. Relays and cooperative diversity, which have not been a huge success in lower frequency cellular networks, may play a more important role in improving coverage in mmWave cellular systems. Many challenges remain in both designing new systems to support mmWave communication and devising algorithms so that mmWave can achieve its best performance in such systems.


The purpose of this article is to provide an overview of the state-of-the-art in signal processing for mmWave wireless communication systems. Section \ref{sec:chan} explains the different channel characteristics at mmWave compared to lower frequency systems. Understanding these characteristics is essential for the design of suitable MIMO architectures and signal processing algorithms. Section \ref{sec:arch} describes the main mmWave MIMO architectures which have been proposed to account for mmWave hardware constraints and channel characteristics. The different approaches described include analog beamforming, hybrid precoding and combining, and one-bit architectures. A detailed review of beamtraining protocols and channel estimation algorithms is provided in Section \ref{sec:train}. Approaches include both codebook-based strategies and compressed channel sensing approaches, and threshold based methods, illustrating approaches that operate under different assumptions. Precoding and combining algorithms for the different mmWave MIMO architectures are introduced in Section \ref{sec:precode}. The objective is to provide some signal processing examples about how MIMO precoders and combiners can be configured in mmWave systems. The paper concludes with some final remarks in Section \ref{sec:conclusion}.


\textbf{Notation}: We use the following notation throughout this paper: bold lowercase $\ba$ is used to denote column vectors, bold uppercase $\bA$ is used to denote matrices, non-bold letters $a,A$ are used to denote scalar values, and caligraphic letters $\cA$ to denote sets. Using this notation, $|a|$ is the magnitude of a scalar, 
$\|\ba\|$ is the  $\ell_2$  norm, $\|\ba\|_{0}$ is the $\ell_0$ norm,  $\|\bA\|_F$ is the Frobenius norm, 
 $\sigma_k(\bA)$ denotes the $k^{th}$ singular value of $\bA$ in decreasing order, $\mathrm{tr}(\bA)$ denotes the trace, 
$\bA^*$ is the conjugate transpose, $\bA^T$ is the matrix transpose,  $\bA^{-1}$ denotes the inverse of a square matrix, 
$[\ba]_k$ is the $k^{th}$ entry of $\ba$, $|\cA|$ is the cardinality of set $\cA$. 
$\bA \otimes \bB$  is the Kronecker product of $\bA$ and $\bB$. We use the notation $\cN(\bm,\bR)$ to denote a complex circularly symmetric Gaussian random vector with mean $\bm$ and covariance $\bR$. We use $\bbE$ to denote expectation.

\section{Millimeter wave propagation and channel models} \label{sec:chan}

Propagation aspects are unique at mmWave due to the very small wavelength compared to the size of most of the objects in the environment.
Understanding these channel characteristics is fundamental to developing signal processing algorithms for mmWave transmitter and receivers.

\subsection{Distance-based path loss}

For free-space propagation,
the transmit power, $P_\text{t}$, and far-field receive power, $P_\text{r}$,
are related by Friis' Law \cite{Rappaport:02},
\begin{equation} \label{eq:Friis}
    P_\text{r} = G_\text{r}G_\text{t}\left(\frac{\lambda}{4\pi d} \right)^2P_\text{t},
\end{equation}
where the powers are in linear scale,
$d$ is the TX-RX separation distance, $\lambda$ is the wavelength
and $G_\text{t}$ and $G_\text{r}$ are the transmit and receive antenna gains.
Friis' Law implies that the \emph{isotropic} path loss
(i.e.\
the ratio $P_\text{t}/P_\text{r}$ with unity antenna gains $G_\text{r}=G_\text{t}=1$),
increases inversely with the wavelength squared, $\lambda^{-2}$.
This fact implies that, in absence of directional antenna gains,
mmWave propagation will experience a higher
path loss relative
to conventional lower frequencies.  For a given
physical antenna aperture, however, the maximum directional gains generally
scale as
$G_\text{r},G_\text{t}\propto \lambda^{-2}$, since more antenna elements can be fit
into the same physical area.  Therefore, the scaling of the antenna gains more than compensates for the increased free-space path-loss at mmWave frequencies.  
Compensating for path loss in this manner will require, however,  directional transmissions with high-dimensional antenna arrays -- explaining how MIMO is  a defining characteristic of
mmWave communication.

While free space propagation can be predicted by Friis' Law,
the path loss in general environments depend on the particular position of objects
that can attenuate, diffract and reflect signals.
Ray tracing has been reasonably successful in predicting site-specific
mmWave propagation, particularly in indoor settings, for at least a decade
\cite{williamson1997investigating,neekzad2007comparison}.  There is also a large body
of work in developing mmWave statistical models that describe the distribution of
path losses  over an ensemble of environments
\cite{tolbert1966attenuation,tharek1988propagation},
with a particularly large number of studies in short-range links
in wireless PAN or indoor LAN systems~\cite{Zwick05,Giannetti:99,Anderson04,Smulders,Manabe,ben2011millimeter,ted2}.
The most common statistical model describes the average path loss (not including small-scale fading)
via a linear model of the form
\begin{equation} \label{eq:PLd}
    PL(d)~[dB] = \alpha + 10\beta\log_{10}(d) + \xi, \quad \xi \sim \cN(0,\sigma^2),
\end{equation}
where $d$ is the distance, $\alpha$ and $\beta$ are linear model parameters and $\xi$ is
 a lognormal term accounting for variances in shadowing.
When converting to dB scale, Friis' formula \eqref{eq:Friis} is
a special case of the model \eqref{eq:PLd} with $\beta = 2$.
Parameters for the model \eqref{eq:PLd} can be found in
~\cite{Zwick05,Giannetti:99,Anderson04,Smulders,Manabe,ben2011millimeter,ted2} for short-range and indoor settings.

More recent work has focused on path loss models for
longer range outdoor links to
assess the feasibility of mmWave picocellular networks,
including measurements in New York City \cite{Rappaport_Access13,rappaport2013broadband,azar201328}.
A surprising consequence of these studies is that,
for distances of up to 200~m
from a potential low-power base station or
access point (similar to cell radii in current micro- and pico-cellular
deployments), the distance-based path loss in mmWave links is no worse than
conventional cellular frequencies after compensating for the additional beamforming gain.
It was these findings that suggested the mmWave bands may be viable for
picocellular deployments and generated considerable interest
in mmWave cellular systems~\cite{Pi_Zhouyue_MCOM11,heath2014}.
At the same time, the results also show that, should mmWave frequencies
be employed in cellular networks, directional transmissions, adaptive
beamforming, and other MIMO techniques will be of central importance.

\subsection{Blocking and outage}
While the distance-based path loss of mmWave frequencies
can be theoretically compensated by directional transmissions,
a more significant challenge is their severe vulnerability to blockage.
Materials such as brick can attenuate mmWave signals by as much as 40 to
80~dB\cite{Allen:94,Anderson04,Alejos:08,Pi_Zhouyue_MCOM11,Rappaport:28NYCPenetrationLoss}
and the human body itself can result in a 20 to 35~dB loss \cite{LuSCP:12}.
Foliage loss can also be significant
\cite{schwering1988millimeter,comparetto1993impact}. Alternatively, humidity and rain fades, common problems for long range mmWave backhaul links~\cite{liebe1989mpm}, are not an issue in either short-range indoor links or micro-cellular systems \cite{Ted:60Gstate11,Rappaport_Access13} with sub-km link distances.

The human body (depending on the material of the clothing) and most building materials are reflective. This allows them to be important scatterers to enable
coverage via NLOS paths for cellular systems
\cite{ben2011millimeter,Rappaport:28NYCPenetrationLoss}.
For example, measurements in New York City \cite{Rappaport_Access13} confirm
that even in extremely dense urban environments, coverage is possible
up to 200~m from a potential cell site. This is good because diffraction -- a primary means of coverage in sub 6 GHz systems -- is not significant at mmWave frequencies.

To quantify the effect of blocking, cellular system evaluation
can use a two-state model (LOS and NLOS) or a three state model (LOS, NLOS, and signal outage). The probability of a link being in each state is a function of distance. Using the NYC measurements in \cite{Rappaport_Access13},
\cite{Akdeniz:14} fits statistical models for this three state model, similar in form
to some LOS-NLOS probabilities used in 3GPP LOS-NLOS for heterogeneous networks \cite{3GPP36.814}.

Blocking models can also be derived analytically from random shape theory \cite{BaiVazHea:14} or from geographic information \cite{KulkSinAnd14}. Using such models, it is possible to evaluate coverage and capacity in mmWave cellular networks analytically using stochastic geometry \cite{BaiHeath:CoverageRate:2015}.

A major outstanding issue is characterizing the
joint probabilities in outage between
links from different cells, which is critical in assessing the benefits of macro-diversity
\cite{RanRapE:14,Ghosh_MmWave:14}.

\subsection{Spatial characteristics and multipath channel models}
\label{sec:chan_mp}

The mmWave MIMO channel can be described with standard multipath models
used in lower frequencies~\cite{TseV:07}.
Consider a MIMO system with $\Nt$ transmit and $\Nr$ receive antennas.
For 2D channel models,
the transmit and receive antenna arrays are described by their
\emph{array steering vectors},
$\aT(\theta_\text{T})$ and $\aR(\theta_\text{R})$
representing the array phase profile as a function of angular directions $\theta_\text{R}$ and $\theta_\text{T}$
of arriving or departing
plane waves.  For an $N$-element uniform linear array (ULA),
the steering vector is given by
\begin{equation}
\ba(\theta) = \left [1, e^{-j2\pi \vartheta}, e^{-j4\pi \vartheta}, \cdots, e^{-j2\pi \vartheta(N-1)} \right ]^T \
\label{steering_vector}
\end{equation}
where the normalized spatial angle $\vartheta$ is related to the physical angle (of arrival or departure) $\theta \in [-\pi/2,\pi/2]$ as $\vartheta = \frac{d}{\lambda} \sin(\theta)$,
$d$ denotes the antenna spacing and $\lambda$ denotes the wavelength of operation.  Typically, $d=\lambda/2$.
In 3D channel models --- which are critical for mmWave arrays ---
the steering vectors are functions $\ba(\theta,\phi)=\ba_\text{az}(\theta)\otimes\ba_\text{el}(\phi)$ of both the
horizontal (azimuth) angle $\theta$ and elevation angle $\phi$ (with
the corresponding normalized elevation angle denoted by $\varphi$).
Given the steering vectors, the MIMO channel can be described by a
multi-path model (see, e.g, \cite{sayeed:02a,sayeed:handbook08,TseV:07})
of the form
\begin{align}
    \by(t) & = \sum_{\ell=1}^{\Np} \alpha_\ell e^{j2\pi  \nu_\ell t}\aR(\theta_{\text{R},\ell},\phi_{\text{R},\ell})
    \aT^*(\theta_{\text{T},\ell},\phi_{\text{T},\ell})\bx(t-\tau_\ell) \nonumber \\
    &  + \bv(t),
\label{sys_mp_td}
\end{align}
where $\bx(t)$ is the transmitted signal vector, $\by(t)$ is the received signal vector,  $\bv(t)$
is the noise vector, and $\Np$ is the number of paths.  Each path $\ell$ is described
by five parameters:  Its angle of arrival pair $(\theta_{\text{R},\ell},\phi_{\text{R},\ell})$,
angle of departure pair $(\theta_{\text{T},\ell},\phi_{\text{T},\ell})$, delay $\tau_\ell$,
complex gain  $\alpha_\ell$ and Doppler shift $\nu_\ell$.  The Doppler shift is determined by the angle of arrival
or departure relative to the motion of the receiver or transmitter.

It is often useful to represent the channel in the frequency domain. In general, the channel response is time-varying
\begin{equation}
    \bH(t,f) = \sum_{\ell=1}^{\Np} \alpha_\ell
        e^{j2\pi (\nu_\ell t - \tau_\ell f)}\aR(\theta_{\text{R},\ell},\phi_{\text{R},\ell})
        \aT^*(\theta_{\text{T},\ell},\phi_{\text{T},\ell}).
    \label{h_wb_mp}
\end{equation}
Suppose that the channel is sufficiently slowly varying over the sigal duration of interest $T$, that is, the Doppler shifts of all the paths are small, $\nu_{\ell}T \ll 1\; \forall \ell, \ell=1,\ldots,\Np$. Then, (\ref{h_wb_mp}) can approximately be expressed as
\begin{equation}
    \bH(f) = \sum_{\ell=1}^{\Np} \alpha_\ell
        e^{-j2\pi \tau_\ell f}\aR(\theta_{\text{R},\ell},\phi_{\text{R},\ell})
        \aT^*(\theta_{\text{T},\ell},\phi_{\text{T},\ell}).
    \label{h_td}
\end{equation}


If in addition, the bandwidth of the channel $W$ is sufficiently
small so that $\tau_\ell W \ll 1 \; \forall \ell, \ell=1,\ldots,\Np$ then we get the narrowband
spatial model for the channel matrix

\begin{equation}
    \bH= \sum_{\ell=1}^{\Np} \alpha_\ell
        \aR(\theta_{\text{R},\ell},\phi_{\text{R},\ell})
        \aT^*(\theta_{\text{T},\ell},\phi_{\text{T},\ell}).
    \label{h_narrowband}
\end{equation}

Statistical MIMO models used for system simulation
typically describe the paths as arriving in ``clusters",
where each cluster has some distribution on the delay, power, and
central angles of arrival and departure.
Physically, the path clusters correspond to different macro-level paths,
and the angle and delay spreads within each cluster capture the scattering from
diffuse reflections along those paths.  MmWave indoor measurements
such as \cite{geng2009millimeter,ted2} have demonstrated large numbers of such path clusters
due to reflections from office materials.  Measurements in New York City
\cite{Rappaport_Access13} have shown that NLOS outdoor links can similarly exhibit several
dominant clusters.  The parameters for statistical multipath models derived from such measurements
can be found in \cite{maltsev2010channel} for 802.11ad systems,
and \cite{Akdeniz:14}, which uses the measurements in \cite{Rappaport_Access13} to
derive statistical multipath models similar to the 3GPP cellular models in
\cite{3GPP36.814,ITU-M.2134}.

 While the above models describe the average statistics of the 
path loss, one major outstanding issue is the modeling of channel variability.
Since mmWave signals can be blocked by many materials, the path clusters can
rapidly appear and disappear, with significant impact on channel tracking.
Some initial stochastic models for temporal variability have appeared in
\cite{jacob2010dynamic}.

\subsection{Beamspace (virtual) system representation}
\label{sec:chan_bs}
The highly directional nature of propagation and the high dimensionality of MIMO channels at mmWave frequencies makes beamspace representation of MIMO systems a natural choice.
The antenna space and beamspace are related through a spatial Fourier transform \cite{sayeed:02a,sayeed:handbook08,sayeed:all10,brady:taps12}. We describe the beamspace representation of a 1D array consisting  of an $N$ dimensional ULA (extension to 2D arrays are straightforward; see, e.g. \cite{sayeed:all10,brady:mu14}). The beamspace (virtual) representation corresponds to system representation with respect to uniformly spaced spatial angles $\vartheta_i = i \Delta \vartheta = i/N$, $i=0, \cdots, N-1$. The corresponding steering vectors defined by $\{ \theta_i = \arcsin(\lambda \vartheta_i/d) \}$ result in an orthonormal basis for the spatial signal space. In particular, the $N \times N$ matrix
\begin{equation}
\bU =\frac{1}{\sqrt{N}} [ \ba(\theta_0), \cdots, \ba(\theta_1), \cdots, \ba(\theta_{N-1}) ]^T \label{U}
\end{equation}
is a unitary DFT matrix: $\bU^* \bU = \bU\bU^* = \bI$.  The
beamspace system representation is given by
\begin{align}
&\bY_\text{b}(f) \approx \bHb(t,f) \bX_\text{b}(f) + \bV_\text{b}(f)    \nonumber\\
&\by_\text{b}(t) = \bU_\text{R}^* \by(t) \ ; \  \bx_\text{b}(t) = \bU_{\text{T}}^*\bx(t) \ ; \   \bv_\text{b}(t) = \bU_\text{R}^* \bv(t)  \label{sys_bs}
\\
&\bHb(t,f) = \bU_\text{R}^* \bH(t,f) \bU_\text{T} \ . \nonumber
\end{align}
which is unitarily equivalent to the antenna domain representation using the transfer function in \eqref{h_wb_mp}. In particular, the sparse/low-rank nature of the MIMO channel at mmWave is explicitly reflected in the sparse nature of the beamspace channel matrix $\bHb(t,f)$.

For a narrowband MIMO system, the beamspace channel representation can be explicitly expressed as \cite{sayeed:02a,Bajwa2010}
\begin{equation}
\bH = \bU_\text{R} \bHb \bU_\text{T}^* = \sum_{i=1}^{\Nr} \sum_{k=1}^{\Nt} [\bHb]_{i,k} \aR(\theta_{\text{R},i})\aT^*(\theta_{\text{T},k})
\label{Hb_nb}
\end{equation}
where $\{\theta_{\text{R},i}\}$ and $\{\theta_{\text{T},k}\}$ are virtual AoAs and AoDs corresponding to the uniformly spaced normalized angles $\{ \vartheta_{\text{R},i} \}$ and $\{ \vartheta_{\text{T},k} \}$.
The concept of beamspace channel representation is intuitive and easy to understand for the narrowband case. It can be extended to time- and frequency-selective channels as well via uniform sampling in delay and Doppler commensurate with the signaling bandwidth $W$ and duration $T$ \cite{sayeed:handbook08,Bajwa2010}:
\begin{align}
\bH(t,f)& =\sum_{i=1}^{\Nr}\sum_{k=1}^{\Nt} \sum_{\ell=0}^{L-1} \sum_{m=-\frac{M}{2}}^{\frac{M}{2}} H_\text{b}(i,k,\ell,m) \aR(\theta_{\text{R},i}) \aT^*(\theta_{\text{T},k}) \nonumber \\
& \times e^{j 2\pi \frac{m}{T}t} e^{-j2\pi \frac{\ell}{W}f}, \label{Hb_tf} \\
\bH(f) & = \sum_{i=1}^{\Nr}\sum_{k=1}^{\Nt} \sum_{\ell=0}^{L-1} H_\text{b}(i,k,\ell) \aR(\theta_{\text{R},i}) \aT^*(\theta_{\text{T},k}) e^{-j2\pi \frac{\ell}{W}f} \label{Hb_f},
\end{align}
where rather than the actual physical delay and Doppler shifts, the channel is represented by uniformly spaced delays $\tau_\ell = \ell/W$ and Doppler shifts $\nu_m = m/T$ with spacings $\Delta \tau = 1/W$ and $\Delta \nu = 1/T$.
$L=\lceil W\tau_{\max} \rceil+1$ and $M=\lceil T\nu_{\max}\rceil$. We note that due to critical sampling in angle, delay, and Doppler, the channel representations in (\ref{Hb_nb}), (\ref{Hb_tf}), and (\ref{Hb_f}) represent multi-dimensional Fourier series expansions with respect to orthogonal Fourier basis functions in angle, delay, Doppler \cite{sayeed:handbook08}.

The wideband channel model needs to be further extended if the number of antennas and/or the signal bandwidth becomes sufficiently large \cite{brady:icc15}. For wideband operation, in general, the spatial angles $\theta_{\text{R},\ell}$ and $\theta_{\text{T},\ell}$ in the arguments of the steering vectors also include a frequency dependence called beam-squint, that can result in significant degradation in performance \cite{60GHz_beamsquint,brady:icc15}. Beam squint is a significant problem for paths for which the dispersion factor $N \alpha \theta_\ell \geq 0.2$ (as applied to the transmit or receive side). A simple multi-beam solution to the beamsquint problem is proposed in \cite{brady:icc15}. If this dispersion factor is sufficiently small for all angles within the angular spread, then the frequency dependence of $\theta(f)$ can be ignored.

\subsection{Beamspace channel sparsity: Low-dimensional communication subspace}
\label{sec:beam_masks}
Consider a channel that is non-selective in time and frequency, $\bH(t,f) \approx \bH$, to focus on its spatial structure.
Let $\sigma_{c}^2=  \tr(\bH^* \bH) = \tr(\bHb^*\bHb) = \sum_{\ell,m} |[\bHb]_{\ell,m}|^2$ denote the channel power. For a given channel realization, the low-dimensional communication subspace is captured by the SVD of $\bH = \bU \mathbf{\Sigma} \bV^*$
We define the effective channel rank, $\peff$, as the number of singular values that capture most of channel power: $ \sum_{i=1}^{\peff} \sigma_i^2(\bH) \geq \eta \sigma^2_c$, for some $\eta$ close to 1 (e.g., 0.8 or 0.9). Optimal communication over the $\peff$-dimensional communication subspace is achieved through the corresponding singular vectors in $\bV$ and $\bU$.

In sparse beamspace MIMO channels, the low-dimensional communication subspace is accessed through Fourier basis vectors that serve as approximate singular vectors for the spatial signal space \cite{sayeed:all10,song:icassp13,brady:taps12,raghavan_sp_mimo:it11}. The channel power is concentrated in a low-dimensional sub-matrix of $\bHb$, denoted ${\tilde \bH}_\text{b}$, consisting of dominant entries indexed by the {\em channel beam masks}:
\begin{equation}
\begin{split}
&\cM = \{ (\ell,m): |[\bHb]_{\ell,m}|^2 \geq \gamma \max_{(\ell,m)} |[\bHb]_{\ell,m}|^2 \}  \ ; \\
&\cM_{\text{r}} = \{ \ell: (\ell,m) \in \cM \} \ , \ \cM_{\text{t}} = \{ m: (\ell,m) \in \cM) \} \ ,
\end{split}
\label{masks}
\end{equation}
where $\gamma \in (0,1)$ is a threshold, $\cM$ is the channel beam mask, and $\cM_{\text{t}}$ and $\cM_{\text{r}}$ denote the transmit and receive masks of dominant beams. The sub-matrix ${\tilde \bH}_{\text{b}}$ is then defined as:
${\tilde \bH}_\text{b} = \left[ [\bHb]]_{\ell,m} \right ]_{\ell \in \cM_{r}, m \in \cM_{\text{t}}}$.
The low-complexity beamspace MIMO transceivers access the low-dimensional communication subspace by selecting the $|\cM_{\text{t}}| \ll \Nt$ transmit beams in $\cM_{\text{t}}$ and $|\cM_{\text{r}}| \ll \Nr$ receive beams in $\cM_{\text{r}}$. We note that $\min(|\cM_{\text{t}}|,|\cM_{\text{r}}|)\approx \peff$ and the performance of these low-dimensional transceivers can be made arbitrarily close to the optimal SVD-based receiver by choosing the threshhold $\gamma$ in (\ref{masks}) sufficiently small so that ${\tilde \bH}_\text{b}$ captures most of the channel power. This discussion applies to deterministic channels. For random multipath variations, $\cM$, $\cM_{\text{t}}$ and $\cM_{\text{r}}$ can be defined by replacing $|[\bHb]_{\ell,m}|^2$ with $\bbE|[\bHb]_{\ell,m}|^2$.

\subsection{Extended virtual representation for the narrowband channel model}\label{sec:extendedvir}

When  any array geometry is considered we can formulate an alternative beamspace representation of the channel, that we will call {\em extended virtual representation}. It is written in terms of  more general dictionaries instead of the basis functions for the DFT.

Consider the multipath narrowband channel model in (\ref{h_narrowband}).
$\bH$ can be written in a more compact way as
\begin{equation}
\bH=\bA_\text{R}\bHb\bA_\text{T}^*,
\end{equation}
where $\bA_\text{T} \in \bbC^{\Nt\times \Np}$  and $\bA_\text{R} \in \bbC^{\Nr\times \Np}$ contain the array response vectors for the transmitter and receiver respectively, and $\bH_{\text{b}}=\text{diag}(\boldsymbol{\alpha})$, with $\boldsymbol{\alpha}=[\alpha_1,\alpha_2,\ldots,\alpha_{\Np}]$.
If we assume that the AoAs/AoDs are taken from a uniform grid of size $G$, i.e. $\theta_{\text{T},\ell}$, $\theta_{\text{R},\ell} \in \{0,\frac{2\pi}{G},\ldots,\frac{2\pi(G-1)}{G}\}$,  with $G \gg \Np$, we can define the array response matrices,  whose columns are the array response vectors  corresponding to the angles in the grid, as $\bar{\bA}_\text{T}$, $\bar{\bA}_\text{R}$. Using these matrices, $\bH$ can be approximated in terms of a $\Np$-sparse matrix $\tilde{\bH}_\text{b} \in \bbC^{G\times G}$, with $\Np$ non zero elements in the positions corresponding to the AoAs and AoDs
\begin{equation}
\bH = \bar{\bA}_\text{R}\tilde{\bH}_\text{b}\bar{\bA}_\text{T}^*.
\label{eq:extendedvirtual}
\end{equation}
There is grid error in \eqref{eq:extendedvirtual}, since the DoAs/DoDs do not necessarily fall to the uniform grid. If the grid size is large enough this error is usually neglected.


\section{MIMO architectures for mmWave communications} \label{sec:arch}
%
%
%
%
%
%
%
%

MIMO technology has already been standardized and is widely used in current commercial WLAN (IEEE 802.11n/ac) and cellular (IEEE 802.16e/m, 3GPP cellular LTE, and LTE Advanced) systems at sub-$6$GHz frequencies \cite{Kim2015,Li2010}.
These standards support a small number of antennas (up to a maximum of eight, although two is commonly used).
The arrays used at mmWave tend to have more elements than lower frequency systems (32 to 256 elements are common), but still occupy a small physical size due to the small wavelength.

\begin{figure}[htb]
\centerline{\includegraphics[width=\columnwidth]{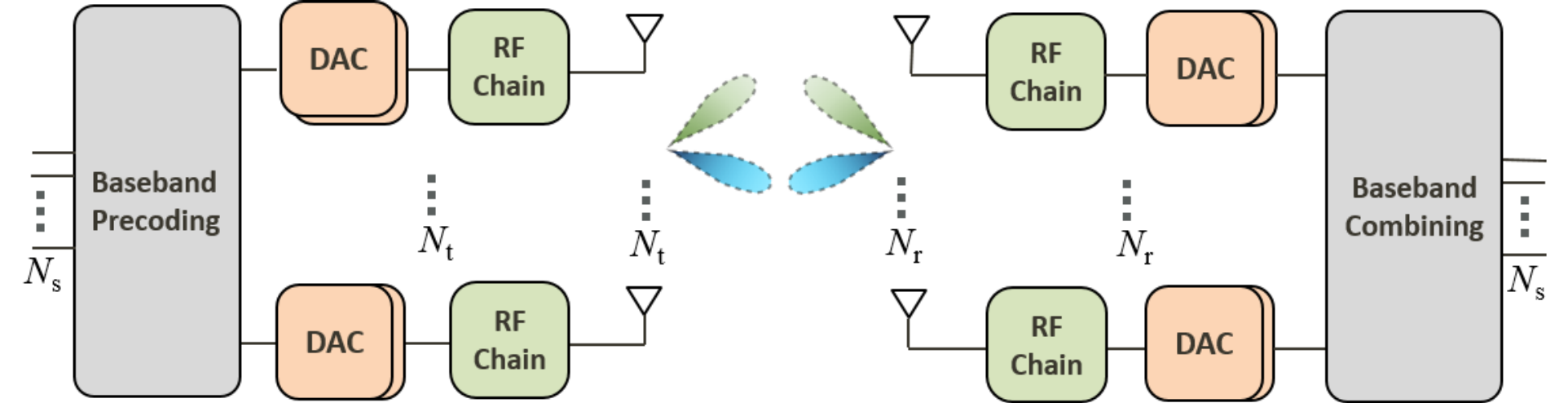}}
\caption{Conventional MIMO architecture at frequencies below $6$GHz.}
\label{fig:conventionalMIMO}
\end{figure}

There are important architectural differences between MIMO communication at sub-$6$GHz frequencies and at mmWave frequencies. At lower frequencies, all the signal processing action happens in the baseband, as illustrated in Fig.~\ref{fig:conventionalMIMO}. Essentially, MIMO at conventional frequencies is an exercise in digital signal processing. 
At higher carrier frequencies and signal bandwidths, there are several hardware constraints  that  make it difficult to have a separate RF chain and data converter for each antenna. 
First, the practical implementation of the
power amplifier (PA) or the low noise amplifier (LNA), the RF chain associated
with each antenna element and all baseband connections is very difficult at mmWave
 \cite{Zhang2015,Doan2004};
these devices have to be packed behind each antenna, and all the antenna elements are placed very close to each other to avoid granting lobes; this space limitation prevents from
using a complete RF chain per antenna. Second, power consumption is also a limiting factor:  (i) PA, ADCs or data interface cards connecting   digital components to DAC/ADCs and are power hungry devices
especially at mmWave \cite{Le2005,Ted:60Gstate11,RapHeaBook}; (ii) a
digital conversion stage per antenna leads to a large
demand on digital signal processing, since many parallel gigasamples per second data streams
have to be proceessed, with an excessive power
consumption as well \cite{Do-Hong2004}.


The exact power consumption depends on the specifications  and technology used to implement a given device.  Table~\ref{power_consumption} shows the range of the power consumed by  different devices included in a mmWave front-end. Data were taken from a number of recent papers proposing protoype devices for  PAs \cite{Floyd2005, LaRocca2009,Yao2006,Dawn2008}, LNAs \cite{Kraemer2009, Fonte2011, Liu2011,Chang2012}, phase shifters \cite{Kim2010,Yu2010,Natarajan2011,Kim2012,Kuo2012,Uzunkol2012,Li2013}, VCOs \cite{Borremans2008,LiuVCO2008,ChenVCO2009} and ADCs \cite{Shettigar2012,SLee2014,Sung2015,Dortz2014,Dong2014,Janssen2013,Miyahara2014} at mmWave frequencies. $\Lt(\Lr)$ is the number of RF chains at the TX(RX).
A detailed treatment of mmWave RF and analog devices and multi-gbps digital baseband circuits can be found in \cite{RapHeaBook}.

\begin{table}[htbp]
\begin{center}
\begin{tabular}{|l|c|c|c|c|}
\hline
{\bf Device} & {\bf Number of devices} & {\bf Power (mW)} \\
& & (single device) \\ \hline \hline
PA & $\Nt(\Nr)$ & 40-250 \\ \hline
LNA & $\Nt(\Nr)$ & 4-86 \\ \hline 
Phase shifter &   $\Nt(\Nr) \times \Lt(\Lr)$ & 15-110  \\ \hline 
ADC &   $\Lt(\Lr)$  &  15-795 \\ \hline 
VCO &  $\Lt(\Lr)$  & 4-25 \\ \hline 
\end{tabular}
\caption{Range for the power consumption for the different devices in a mmWave front-end.}
\label{power_consumption}
\end{center}
\end{table}

The hardware constraints have led to several mmWave-specific MIMO architectures where signal processing is accomplished in a mixture of analog and digital domains or where different design tradeoffs are made with respect to number of antennas or resolution of data converters. This section reviews several MIMO architectures for mmWave systems, highlighting the implications on signal processing. 

There are other implications of different hardware, beyond what is considered in this section, where signal processing can play a role. Examples include phase noise \cite{Bourdoux2006,Choi2006,Erceg2009}, IQ imbalance \cite{Rizvi2008, Gomes2014}, and nonlinear effects \cite{Schenk2008,Choi2006,Erceg2009}. Modeling these impairments and developing digital compensation algorithms for mmWave is an active area of research \cite{Zhang2013,Fan2014,Suyama2012}.

\subsection{Analog beamforming}

\begin{figure}[htb]
\centerline{\includegraphics[width=\columnwidth]{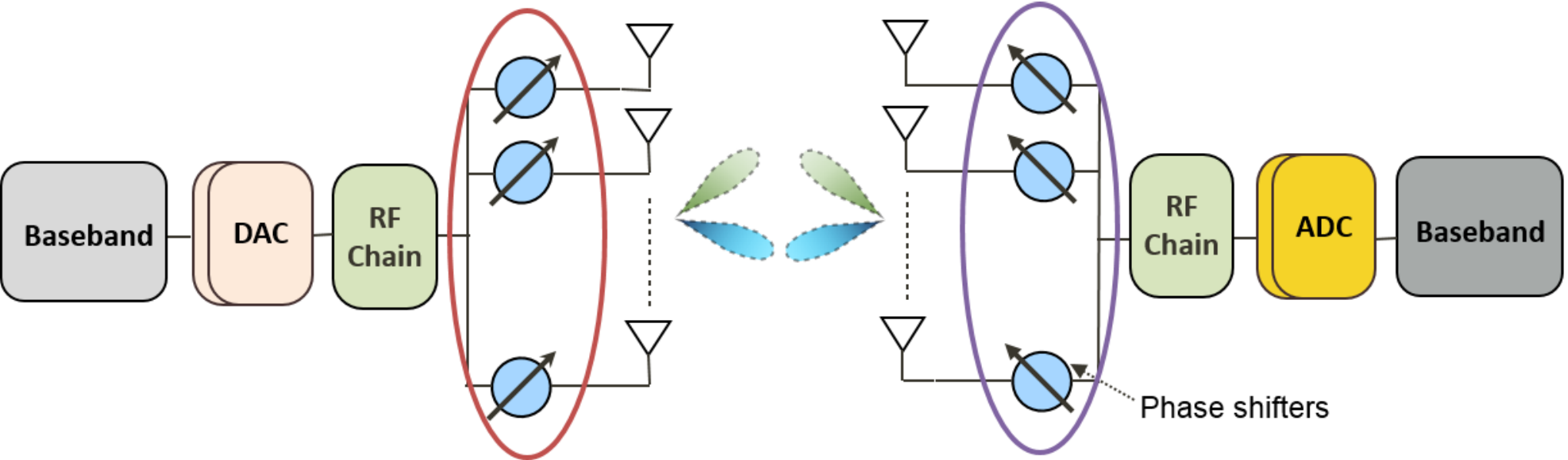}}
\caption{MmWave MIMO system using analog only beamforming.}
\label{fig:analogBF}
\end{figure}

Analog beamforming is one of the simplest approaches for applying MIMO in mmWave systems. It can be applied at both the transmitter and receiver. It is defacto solution supported in IEEE 802.11ad \cite{80211ad}. 

Analog beamforming is often implemented using a network of digitally controlled phase shifters. In this configuration, several antenna elements are connected via phase shifters to a single RF chain, as illustrated in Fig.~\ref{fig:analogBF}. Other configurations are possible where the combining happens at an intermediate frequency \cite{Fakharzadeh2010}. The phase shifter weights are adaptively adjusted using digital signal processing using a specific strategy to steer the beam and meet a given objective, for example to maximize received signal power. 

The performance achieved with analog beamforming based on phased arrays is limited by the use of quantized phase shifts and the lack of amplitude adjustment. This makes it more challenging to finely tune the beams and steer nulls.  RF phase shifters may be active or passive. Practical active phase shifters also introduce performance degradation  due to phase-shifter loss, noise and non linearity.  Although passive phase shifters have a lower consumption and do not introduce non-linear distortion, they  occupy a larger area and incur larger insertion losses \cite{Poon2012}. The power consumed by the phase shifters also depends on the resolution of the quantized phases. 

There are several implications of using analog beamforming for mmWave MIMO. Analog beamforming with a single beamformer only supports single-user and single-stream transmission. This means it is not possible to realize multi-stream or multi-user benefits associated with MIMO. Steering the beams is not trivial, especially when a communication link has not yet been established. This leads to the need for beam training algorithms (described in Section \ref{sec:beamtrain}) and techniques for channel estimation (described in Section \ref{sec:train}). In general, to achieve the highest performance, the wireless protocol should be designed to support beam steering \cite{Wang2009}.
 
\subsection{Hybrid analog-digital precoding and combining}
 
\begin{figure}[htb]
\centerline{\includegraphics[width=1\columnwidth]{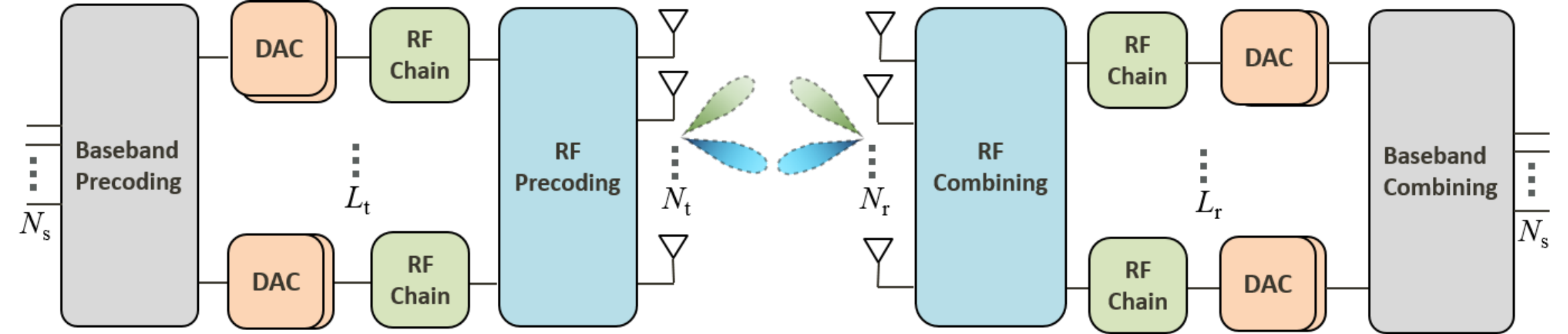}}
\caption{MIMO architecture at mmWave based on hybrid analog-digital precoding and combining.}
\label{fig:arch-hybrid}
\end{figure}

Hybrid architectures are one approach for providing enhanced benefits of MIMO communication at mmWave frequencies. This architecture, shown in Fig.~\ref{fig:arch-hybrid}, divides the  MIMO optimization process between analog and digital domains. A small number of transceivers are assumed (2 to 8), so that $\Ns<\Lt<\Nt$ and $\Nr>\Lr>\Ns$. Assuming that $\Ns>1$, then the hybrid approach allows spatial multiplexing and multiuser MIMO to be implemented; analog beamforming is a special case when $\Ns=\Lt=\Lr=1$. WirelessHD described the application of a hybrid architecture \cite{WirelessHD}, but to our knowledge it has not yet been commercialized. 
Hybrid architectures were investigated at lower frequencies in \cite{Molisch2005,Sudarshan2006,Venkateswaran2010}. The general concept of hybrid precoding introduced in this prior work  can also be applied to mmWave systems. The algorithms for the design of the precoders/combiners described in these papers use 
however channel models that do not fully capture the effect of limited mmWave scattering and large arrays. While those algorithms  can be used at mmWave frequencies, further simplifications occur when the sparsity of the mmWave channel is leveraged. A comparison of performance and complexity of specific  mmWave hybrid precoding schemes and general hybrid precoding algorithms is a topic of current research.

\begin{figure}[htb]
\begin{center}
\begin{tabular}{cc}
\hspace*{-5mm}
\includegraphics[width=0.25\columnwidth]{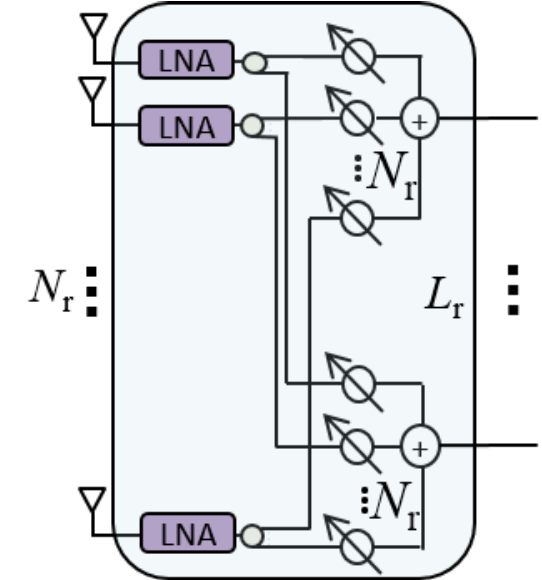}
&
\hspace*{5mm}
\includegraphics[width=0.25\columnwidth]{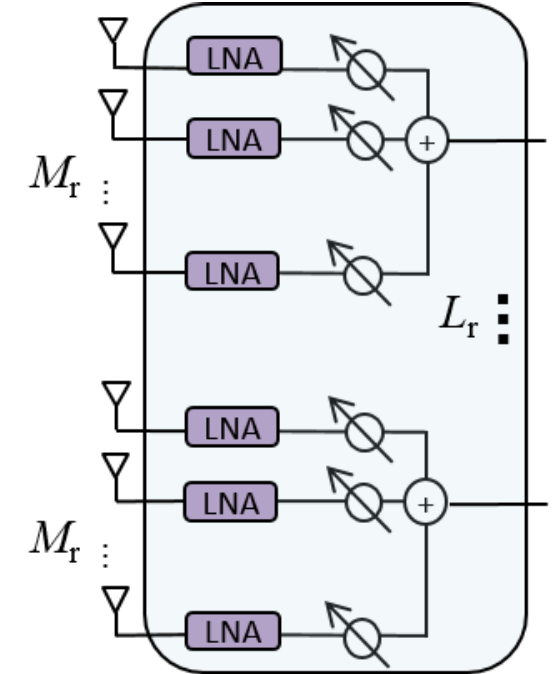}\\
\hspace*{3mm}(a) & \hspace*{17mm} (b)\\
\end{tabular}
\end{center}
\caption{Analog processing for hybrid beamforming based on phase shifters: (a) each RF chain is connected to all the antennas; (b) each RF chain is connected to a subset of antennas. }
\label{fig:phase-shifters}
\end{figure}

The RF precoding/combining stage can be implemented using different analog approaches like phase shifters \cite{Hajimiri2005,Pivit2013}, switches \cite{Wang2008} or lenses. Two hybrid structures are possible \cite{Han2015}. In the first one, all the antennas can connect to each RF chain (as illustrated in Fig.~\ref{fig:phase-shifters}(a)). In the second one (see  Fig.~\ref{fig:phase-shifters}(b)), the array can be divided into subarrays, where each subarray connects to its own individual transceiver. Having multiple subarrays reduces hardware complexity at the expense of less overall array flexibility. A complete analysis of the energy efficiency and spectrum-efficieny of both architectures is provided in \cite{Han2015}. Massive hybrid architectures based on the subarray structure are analyzed in \cite{Zhang2015}. Some prototypes for hybrid mmWave MIMO systems are also being developed \cite{Roh2014, Cudak2014,Pisek2014}.

Two different realizations of the hybrid architecture are illustrated in Fig.~\ref{fig:phase-shifters}. A hybrid precoder/combiner based on phase shifters would normally use digitally controlled phase shifters with a small number of quantized phases. An advantage of the hybrid approach is that the digital precoder/combiner can correct for lack of precision in the analog, for example to cancel residual multi-stream interference. This allows hybrid precoding to approach the performance of the unconstrained solutions \cite{ElAyach_2014,Alkhateeb_JSTSP14}.
Hybrid precoding is a topic of substantial current research \cite{AMGH2014,Liang2014,Kim2013,Kim2014,Zhang2014,Chen2015}.

\begin{figure}[htb]
\begin{center}
\begin{tabular}{cc}
\hspace*{-5mm}\includegraphics[width=0.2\columnwidth]{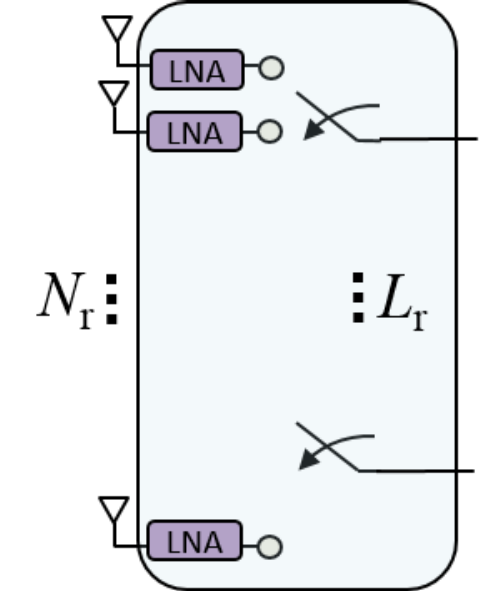}
&
\hspace*{5mm}\includegraphics[width=0.2\columnwidth]{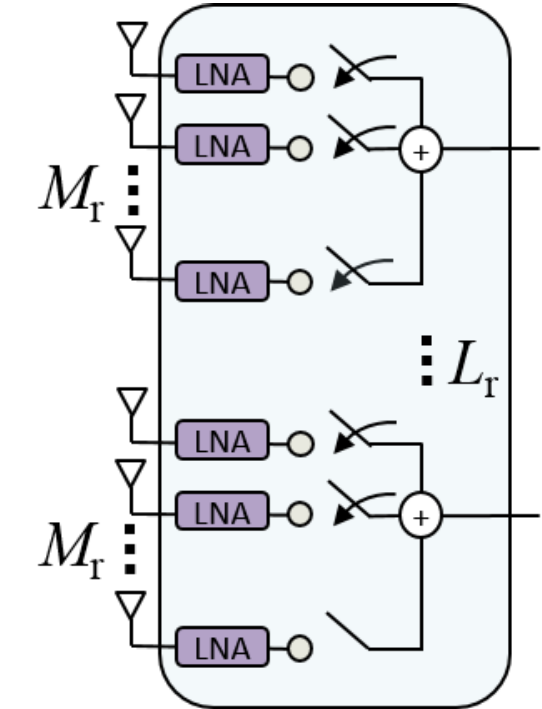}\\
(a) & \hspace*{12mm}(b)\\
\end{tabular}
\end{center}
\caption{Analog processing for hybrid beamforming based on switches: (a) each RF chain can be  connected to all the antennas; (b) each RF chain can be connected to a subset of antennas. }
\label{fig:arch-switches}
\end{figure}

An alternative mmWave hybrid architecture that makes use of switching networks \cite{Koch2009, Reyaz2012} with small losses \cite{Wang2008} has been recently proposed \cite{Roi2015}, to further reduce complexity and power consumption of the hybrid architecture based on phase shifters. This architecture, illustrated in Fig.~\ref{fig:arch-switches}, exploits the sparse nature of the mmwave channel by implementing a compressed spatial sampling of the received signal. The analog combiner design is performed by a subset antenna selection algorithm instead of an optimization over all quantized phase values. Every switch can be connected to all the antennas if the array size is small or to a subset of antennas for larger arrays.  

\begin{figure}[htb]
\centerline{\includegraphics[width=0.45\columnwidth]{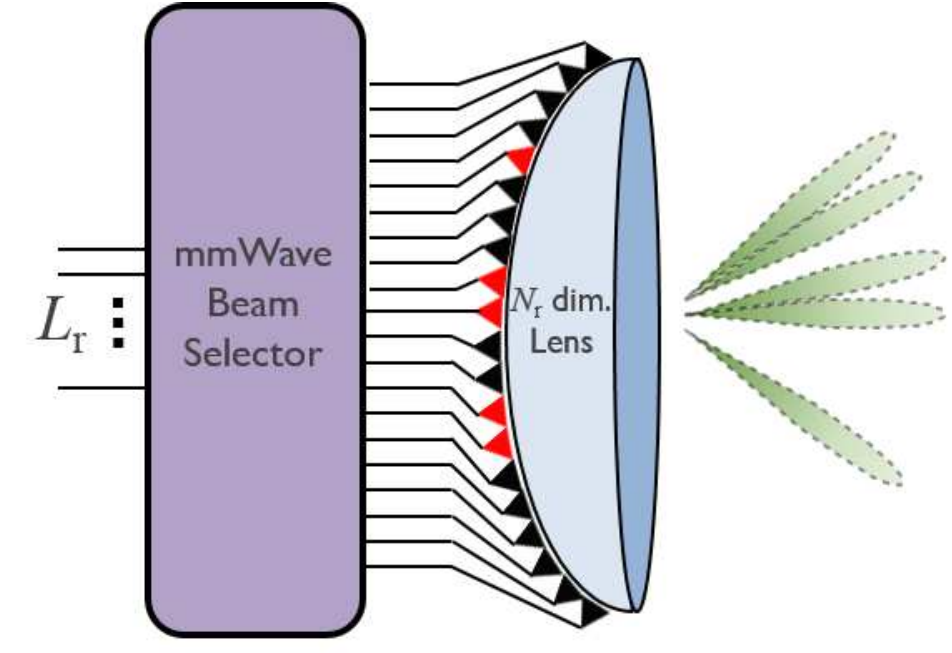}}
\caption{\sl \scriptsize The CAP-MIMO transceiver that uses a lens-based front-end for analog beamforming; it maps the $p = \Ns$ precoded data streams to $L = O(p)$ beams via the mmWave beam selector and lens.}
\label{fig:arch-lenses}
\end{figure}

Analog beamforming for $\Ns>1$ in the hybrid architecture can also be realized using a lens antenna at the front-end, using the fundamental fact that lenses compute a spatial Fourier transform thereby enabling direct channel access in beamspace \cite{sayeed:all10,brady:taps12}. This continuous aperture phased (CAP) MIMO transceiver architecture is illustrated in \fig{fig:arch-lenses} and suggests a practical pathway for realizing high dimensional MIMO transceivers at mmWave frequencies with significantly low hardware complexity compared to conventional approaches based on digital beamforming. The antennas and RF precoder/combiner in \fig{fig:arch-hybrid} are replaced by the continuous-aperture lens antenna and mmWave beam selector in \fig{fig:arch-lenses}.
CAP-MIMO directly samples in beamspace via an array of feed antennas arranged on the focal surface of the lens antenna. 

CAP-MIMO enables direct access to the beamspace channel matrix $\bH_{\text{b}}$; see (\ref{sys_bs}); in particular, lens-based front-end represents an analog realization of the beamforming matrix $\bU$. With a properly designed front-end, different feed antennas excite (approximately) orthogonal spatial beams that span the coverage area \cite{brady:taps12}. The number of ADC/DAC modules and transmit/receive  chains tracks the number of data streams $\Ns=p$, as in the phase-array-based hybrid transceiver, as opposed to the number of antennas $\Nt/\Nr$ in the conventional massive MIMO architecture. However, the mapping of the $\Ns$ (precoded) data streams into corresponding beams is accomplished via the mmWave beam selector that maps the mmWave signal for a particular data stream into a feed antenna representing the corresponding beam. The wideband lens can be designed in a number of efficient ways, including a discrete lens array (DLA) for lower frequencies or a dielectric lens at higher frequencies \cite{brady:taps12}.

There are many implications of using a hybrid architecture for mmWave MIMO. Given channel state information, new algorithms are needed to design the separate precoders/combiners since they decompose into products of matrices with different constraints (see Section \ref{sec:hybridprecode} and Section \ref{sec:analogbeam} for more information). Learning the channel state is also harder, since training data is sent through analog precoders and combiners (see Section \ref{sec:train}). More challenges are found when going to broadband channels as the analog processing is (ideally) frequency flat while the digital processing can be frequency selective. There are many opportunities for future research into designing cellular or local area networks around support for hybrid architectures.

\subsection{Low resolution receivers}

\begin{figure}[htb]
\centerline{\includegraphics[width=0.6\columnwidth]{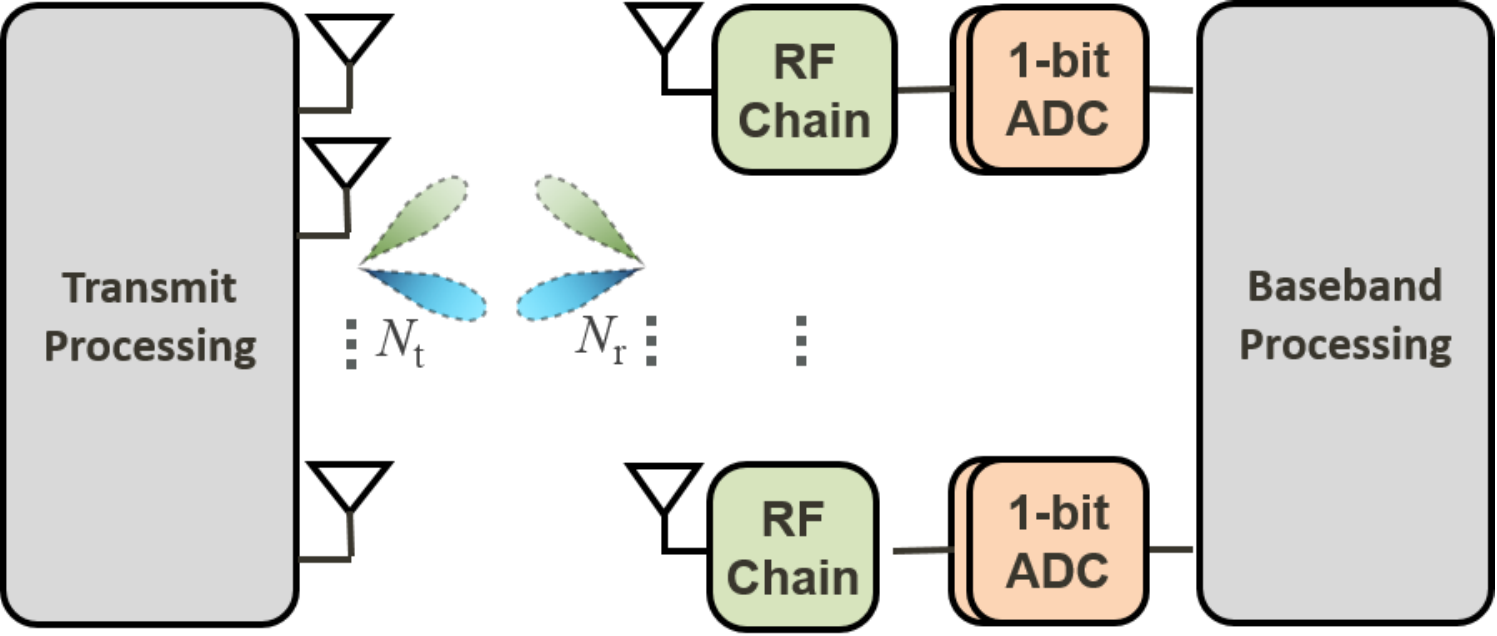}}
\caption{One-bit receiver at mmWave.}
\label{fig:1-bit}
\end{figure}

An alternative to analog and hybrid architectures at the receiver is to reduce the resolution and thus power consumption of the ADCs to a few or as little as one bit. This leads to a different approach as illustrated in Fig.~\ref{fig:1-bit}, where a pair of low resolution ADCs are used to sample the in-phase and quadrature components of the demodulated signal at the output of each RF chain. This makes a tradeoff between having more RF chains and fewer power hungry ADCs. The case of a one-bit ADC is especially interesting as it has negligible power consumption compared to other components in the front-end (a one-bit ADC at ultra-high sampling rate of $240$ GS/s consumes around $10$ mW \cite{FettweisCTW14}). Data interface circuits connecting digital components to DAC/ADCs are also power hungry when working at mmWave frequencies \cite{Doan2004}. The power consumed by the high speed interfacing cards also depends on the resolution, so reducing the number of bits in the ADC not only reduces the power consumed by the front-end in the MIMO receiver, but also limits the consumption of the baseband circuitry.

The fundamentals of communicating with one-bits ADCs are different \cite{Dabeer_SPAWC06, Mezghani_ISIT07, Singh_TCOM09, Mezghani_ISIT12,Mo_Jianhua_arxiv14}. For example, the optimum signal constellation is discrete and is limited by the ADC resolution at the receiver. In MIMO systems, the low SNR capacity gap between one-bit and infinite-resolution ADC is only 1.96 dB \cite{Mezghani_ISIT07}. At high SNR, at most $2^{2 \Nr}$bps/Hz is achievable if the rank of the channel is no less than $\Nr$. Capacity characterization with low-resolution ADCs is an ongoing research challenge. 
 
The use of few- and one-bit ADCs has several signal processing implications. The role of channel state information is different, e.g. channel inversion precoding may be better than eigenbeamforming \cite{Mo_Jianhua_arxiv14}, as discussed further in \secref{sec:onebitcombine}. This might lead to different hybrid precoding optimizations that are compatible with one-bit ADCs. Acquiring channel state information is also more challenging.  Although the channel-estimation error with one-bit ADCs decreases at best quadratically per measurement bit (versus exponentially in the conventional case), it also decreases with the sparsity of the channel \cite{Jacques_IT13}. This suggests that relatively few measurements may suffice and that one-bit compressive sensing algorithms can be employed for channel estimation \cite{OneBitMo2014}, as discussed further in \secref{sec:onebitchanest}. 
Future work is still needed to develop mmWave specific channel estimation algorithms, especially those designed in conjunction with appropriate transmit and receive signal processing algorithms.

\section{Precoding and combining} \label{sec:precode}
Precoding and combining is different at mmWave for three main reasons. 
\begin{enumerate}
\item There are more parameters to configure, due to the different array architectures as described in \secref{sec:train}. This requires different algorithms for finding both the analog and digital parameters, and makes the resulting algorithms architecture-dependent. 
\item The channel is experienced by the receiver through the analog precoding and combining. This means that the channel and the analog beamforming are intertwined, making estimation of the channel directly a challenge. 
\item There is more sparsity and structure in the channel, resulting from the use of large closely spaced arrays and large bandwidths. This provides structure that can be exploited by signal processing algorithms. 
\end{enumerate}
In this section, we describe signal processing techniques for configuring mmWave transmit and receive arrays. We consider approaches that do not use explicit knowledge of the channel (beam training) as well as hybrid precoding / combining techniques that make use of an estimate of the channel, provided by the algorithms developed in \secref{sec:train}.  The algorithms are described using a narrowband channel model. Extensions to frequency selective channels in many cases is still ongoing research, due to the difficulty in implementing adaptive frequency selective filtering in the analog domain. 

%
%
%
%
%
%

\subsection{Beam training protocols} \label{sec:beamtrain}

Analog beamformers in mmWave are usually designed using a {\em closed-loop} beam training strategy, based on using a codebook which includes beam patterns at different resolutions. Some simple protocols use an iterative process to exchange information between the transmitter and receiver using a narrower and narrower beamwidth at each step, with the purpose of discovering the angular directions of the strongest signal between the receiver and transmitter (i.e. the best angle-of-arrival and angle-of-departure), without explicit channel estimation. Codebook beam training strategies \cite{Wang2009,Tsang2011,Zhou2012,Hur2013,Hosoya2014,Alkhateeb2014,Thomas2014}  use an iterative process to measure the angular power over its codebook. Each code in the codebook directs the beam in a particular angular direction.
These training strategies  have been implemented in standards like IEEE 802.15.3c, IEEE 802.11ad, and Wireless HD,  which assume analog-only beamforming and single-stream transmission.

\begin{figure}[htb]
\centerline{\includegraphics[width=0.8\columnwidth]{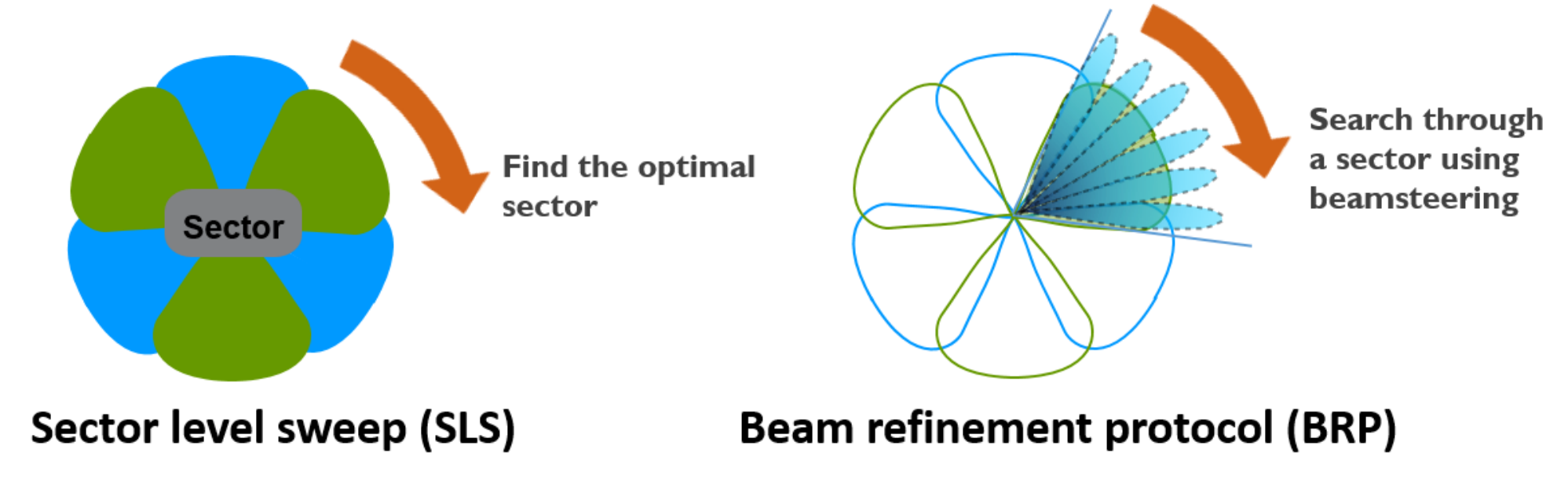}}
\caption{Coarse and fine beam adaptation steps in the 802.11ad beamtraining protocol.}
\label{fig:beamtraining80211ad}
\end{figure}

To illustrate the beam training concept, we describe the protocol in IEEE 802.11ad \cite{80211ad}. It uses a specified period called Beam Training Interval (BTI) for the iterative process to occur. This procedure  includes  three phases illustrated in Fig.~\ref{fig:beamtraining80211ad}: a) {\em Sector  Level Sweep (SLS)}, a coarse beam adaptation which trains a combination of sector (at one end) and antenna (at the other end). The access point transmits the Initiator Transmit Sector Sweep (TXSS) on each of its sectors up to a maximum of 64 sectors per antenna and a total maximum number of sectors of 128. After the access point completes its sweep, the users use carrier sense multiple access (CSMA) with an exponential backoff to begin the Responder Sector Sweep (RSS).  b) {\em Beam Refinement Protocol (BRP)}, a  fine beam training step which can be used to increase the quality of the link if required; it involves training of different transmit and/or receive antenna configurations. BRP packets append special training to IEEE 802.11ad data packets. This training field allows either the receiver or transmitter (but not both at the same time) to try a new antenna beam. The BRP packet includes training for channel estimation with the new antenna beam. If the transmitter is refining its beam, the receiver sends feedback to the transmitter regarding the success or failure of the new beam. c) {\em Beam tracking}, a periodic refinement over a small number of antenna configurations.

A generalization of beam training to hybrid precoding is provided in \cite{Alkhateeb2014}, assuming a hybrid architecture based on phased shifters. This approach uses a beam training process that involves adaptive measurements over a multiresolution dictionary and a low complexity bisection strategy for the sparse recovery. The algorithm estimates  the parameters (AoA/AoD and path gain) of one path per iteration after subtracting the contribution of the previously estimated paths. To estimate each path's parameters, an adaptive search over the AoA/AoD is performed starting with wide beams in the early stages and narrowing the search based on the estimation outputs in the later stages to focus only on the most promising directions.  To implement these adaptive beams, a novel multi-resolution beamforming codebook was also developed. The codebook construction idea depends on approximating the ideal sectored beam patterns directly using hybrid analog/digital precoders.  The main advantage of this hybrid precoding based codebook compared with prior analog-only multi-resolution codebooks is the higher design degrees of freedom given by the extra digital processing layer, which lead to better beam patterns and more flexibility with RF phase shifter limitations.  One drawback of the adaptive scheme in \cite{Alkhateeb2014} is the need for a feedback link between the transmitter and receiver. This has been addressed in \cite{Alkhateeb2014c} where a ping-pong algorithm was used along the same lines of \cite{Alkhateeb2014} to estimate multi-path mmWave channels.

\subsection{Hybrid precoding}\label{sec:hybridprecode}

Hybrid precoding offers a compromise between system performance and hardware complexity. The precoding/combining processing is divided between the analog and digital domains. A number of RF chains much less than the number of antennas is required \cite{Zhang2005a,Venkateswaran2010,Roh2013,ElAyach2013,Alkhateeb2013,ElAyach_2014,Alkhateeb2014,Kim2013,Han2015,Roi2015}.
In this section we review several hybrid precoding/combining strategies for the single-user and multi-user cases and for the different hybrid architectures.

From Fig.~\ref{fig:arch-hybrid}, assuming flat-fading and perfect synchronization, the discrete-time model for the received signal for a single symbol period is 
\begin{equation}
\mathbf{y}=\sqrt{\rho}\mathbf{W}^*\mathbf{H}\mathbf{F}\mathbf{s}+\mathbf{W}^*\mathbf{n},
\end{equation}
where
$\rho$ represents the average transmitted power per symbol, and $\mathbf{n} \in \mathbb{C}^{N_\text{r}\times 1}$ is the noise vector with $\mathcal{N}(0,\sigma_n^2)$ entries. 
$\mathbf{F}=\mathbf{F}_\text{RF} \mathbf{F}_\text{BB}$ is composed of an RF precoder $\mathbf{F}_\text{RF} \in \mathbb{C}^{N_\text{t} \times L_\text{t}}$ and a baseband precoder $\mathbf{F}_\text{BB} \in \mathbb{C}^{L_\text{t} \times N_\text{s}}$.
Equivalently, the hybrid combiner $\mathbf{W}=\mathbf{W}_\text{RF} \mathbf{W}_\text{BB}$ is composed of an RF combiner $\mathbf{W}_\text{RF} \in \mathbb{C}^{N_\text{r} \times L_\text{r}}$, and a baseband combiner $\mathbf{W}_\text{BB} \in \mathbb{C}^{L_\text{r} \times N_\text{s}}$.
The precoding and combining matrices $\mathbf{F}_\text{RF}$ and $\mathbf{W}_\text{RF}$ are subject to specific constraints depending on the hardware architecture for the RF beamforming stage. 

\subsection{Single user hybrid precoding with phase shifters or switches}\label{sec:hybridprecodenolens}

In \cite{Zhang2005a,Venkateswaran2010,ElAyach_2014,ElAyach2013,Alkhateeb2013,Alkhateeb2014,Kim2013,Han2015}, precoding/combining processing is divided between the baseband, which uses digital hardware, and the RF that employs a network of phase shifters. A hybrid system based on phase shifters (see Fig~\ref{fig:phase-shifters}), imposes the constraint of 
unit norm entries in $\mathbf{W}_\text{RF}$ and $\mathbf{F}_\text{RF}$ and further possibly quantized. In \cite{Zhang2005a,Venkateswaran2010}, hybrid analog/digital precoding which does not exploit channel structure  was considered for both spatial diversity and multiplexing systems. Other algorithms have been specifically designed for mmWave systems, leveraging  the special characteristics of mmWave channels to simplify the design. 

A general approach for hybrid precoding would be to maximize the mutual information given by $I(\rho, \mathbf{F}_\text{RF}, \mathbf{F}_\text{BB}, \mathbf{W}_\text{RF}, \mathbf{W}_\text{BB}) $
\begin{align}
= 
\log \left| 
\bI + \rho \bR_{\text{n}}^{-1} \bW^* \bH \bF \bF^* \bH^* \bW 
\right|  \label{eq:entireMutualInfoExpr}
\end{align}
where $\bR_{\text{n}} = \bW^* \bW$ and using the definitions of $\bF$ and $\bW$ from \secref{sec:hybridprecode}. Optimizing \eqref{eq:entireMutualInfoExpr} directly is challenging due to the constraint sets. An alternative proposed in \cite{ElAyach_2014} is to assume that the receiver performs ideal decoding, neglecting the receiver hybrid constraint. Effectively this removes the terms that depend on $\bW$ from \eqref{eq:entireMutualInfoExpr}. With some approximations, this leads to a new problem where the hybrid precoders are found by approximating the unconstrained optimal precoder $\bF_{\text{opt}}$,  given by the channel singular value decomposition (SVD) solution
\begin{eqnarray}
(\bF_{\text{RF}}^{opt},\bF_{\text{BB}}^{opt}) & = & \argmin_{ \bF_{\text{BB}},\bF_{\text{RF}}}\left\lVert \bF_{\text{opt}}- \bF_{\text{BB}}\bF_{\text{RF}}\right\rVert_{F}, \nonumber \\
& & \nonumber \\
& \text{s.t}. &  \bF_{\text{RF}} \in \cF_{\text{RF}}, \nonumber \\
& & \left\lVert \bF_{\text{RF}}\bF_{\text{BB}}\right\rVert^2_{F}=\Ns,
\end{eqnarray}
where $\cF_{\text{RF}}$  is the set of feasible RF precoders which correspond to a hybrid architecture based on phase shifters, i.e., the set of $\Nt \times N_{\text{RF}}$ matrices with constant-magnitude entries. To solve this problem, an orthogonal matching pursuit (OMP) based algorithm was proposed in \cite{ElAyach_2014}. It uses a sparse channel model like in (\ref{eq:extendedvirtual}) and proposes a related problem that involves configuring the RF beamforming vectors from a dictionary of steering vectors based on channel AoDs. This solution was found to be close to the unconstrained digital solution and offer substantial gains over the case of single-stream analog beamforming. The hybrid precoding design problem based on the dictonary approach is extended to an architecture based on subarrays in \cite{Ramakrishna2015}; the sparsity of the channel is also used to define an efficient way to find the near-optimal precoder. In \cite{Rahman2014} the codebook base approach is also considered, and another method for the efficient selection of the precoders/combiners is presented.
In \cite{RusuICC2015}, the semi-unitary structure of the optimum  precoder  (in the absence of hardware constraints) is exploited. The search space in the array manifold  is significantly reduced and a much lower complexity optimization algorithm is obtained. In \cite{Sohrabi2015} the hybrid structure based on phase shifters is further analyzed. 
It is theoretically shown that if $\Lr,\Lt\geq 2\Ns$,  the hybrid system performs 
as the all-digital precoding/combining scheme. This work also proposes an aternative design strategy for the precoders/combiners when $\Lr=\Lt=\Ns$, which performs close to the fully-digital solution.
Another solution presented in \cite{Chen2015} performs a simplex 1-D iterative local search for every
element of the analog precoder;  the large number of entries which are updated separately increases the computational complexity.

The design of combiners when the receiver hybrid architecture is based on switches (see Fig.~\ref{fig:arch-switches}) instead of phase shifters has been addressed in  \cite{Roi2015}. The RF combining/precoding matrices become selection matrices routing $L_\text{r},L_\text{t}$ antennas to the corresponding RF chain.
Each column of $\mathbf{W}_\text{RF},\mathbf{F}_\text{RF}$ is a binary vector with a single one and zeros elsewhere.
The combiner design that maximizes mutual information is a combinatorial problem. After some approximations a sparse reconstruction problem can also be formulated and solved using a variant of simultaneous orthogonal matching pursuit (SOMP). 

Most work on hybrid precoding like \cite{ElAyach_2014,RusuICC2015,Sohrabi2015} requires the availability of channel knowledge, at least at the receiver. To relax this assumption, \cite{Alkhateeb2013} develops a hybrid precoding algorithm for mmWave systems based on partial channel knowledge. With a two-stage algorithm, \cite{Alkhateeb2013} showed that the hybrid precoding performance with perfect channel knowledge can be approached when each of the transmitter and receiver knows only its AoDs (or AoAs). Relaxations for hybrid precoding with no channel knowledge and with quantized phase shifters has been considered in \cite{Alkhateeb2014}. Other extensions are made for single-stream MIMO-OFDM where the analog/digital precoders are designed to maximize either the signal strength or the sum-rate over different sub-carriers \cite{Kim2013}. Other variations of hybrid precoding with arrays of sub-arrays of phase shifters were considered in \cite{ElAyach2013,Han2015}. It was shown here that this system incurs a small loss compared to the fully-connected architecture, while the power consumption is lower. Many other extensions are also important, like hybrid precoding codebook design, and wideband hybrid precoding (see \cite{AMGH2014} for more suggested future work).

\subsection{Single-user hybrid precoding and combining with lens-based front-end} \label{sec:analogbeam}

Precoding and combining for lens-based analog beamforming makes use of the beamspace system representation in (\ref{sys_bs}) to exploit the resulting sparsity in the thresholded  sub-matrix ${\tilde \bH}_{\text{b}}$ defined in \secref{sec:beam_masks}. If CSI is available at the transmitter, an SVD of $\tilde{\bH}_{\text{b}} = \tilde{\bU}_{\text{b}} \tilde{\mathbf{\Sigma}}_{\text{b}} \tilde{\bV}_{\text{b}}^*$ may be used \cite{sayeed:all10} for precoding. The matrix $\tilde{\bV}_{\text{b}}$ is used for precoding at the transmitter and $\tilde{\bU}_{\text{b}}$ is used for post-processing at the receiver to create $p_{\mathrm{eff}} = \min(|\cM_{\text{r}}|,|\cM_{\text{t}}|)$ orthogonal channels. A simpler approach exploits the fact that the Fourier (beamspace) basis vectors serve as approximate eigenvectors for sparse beamspace mmWave MIMO channels. In this case, no precoding is done at the transmitter, except possibly some power allocation across the $p_{\mathrm{eff}}$ transmit data streams. Residual interference between the different data streams is suppressed via post-processing at the receiver, e.g., the MMSE receiver \cite{song:icassp13}. By appropriate thresholding so that most of the channel power is captured by ${\tilde \bH}_{\text{b}}$, both approaches deliver near optimal performance \cite{song:icassp13}.

\subsection{Precoding and combining with 1-bit ADCs} \label{sec:onebitcombine}

In \cite{Mo_Jianhua_arxiv14}, where {CSIT} is assumed, simple channel inversion precoding (versus the usual eigenbeamforming) is shown to be nearly optimal if the channel has full row rank. MIMO precoding eliminates the gap between unquantized and quantized achievable rates at low and medium SNRs, and provides a substantial performance improvement compared with the no precoding case. If full row rank is not true, a different precoding method is proposed achieving the high SNR capacity. Despite this potential gain, limited feedback precoding with 1-bit ADCs, including suitable codebook design, remains as an open problem. Further, most work on low resolution ADCs has focused on the single user MIMO setting, and  there has been limited work on the multiuser case.

\subsection{Multiuser extensions} \label{sec:multiuser} 

Multiuser precoding at mmWave is still an active area of research \cite{brady:gcom13,brady:mu14,Alkhateeb2014a}. The basic idea of most multiuser approaches is to assign different analog beams to different users then possibly use baseband digital processing to further reduce inter-user interference. 

\subsubsection{Multiuser precoding and combining in lens-based hybrid architecture}
\begin{figure}[htb]
\begin{tabular}{c}
\centerline{\includegraphics[width=2.5in]{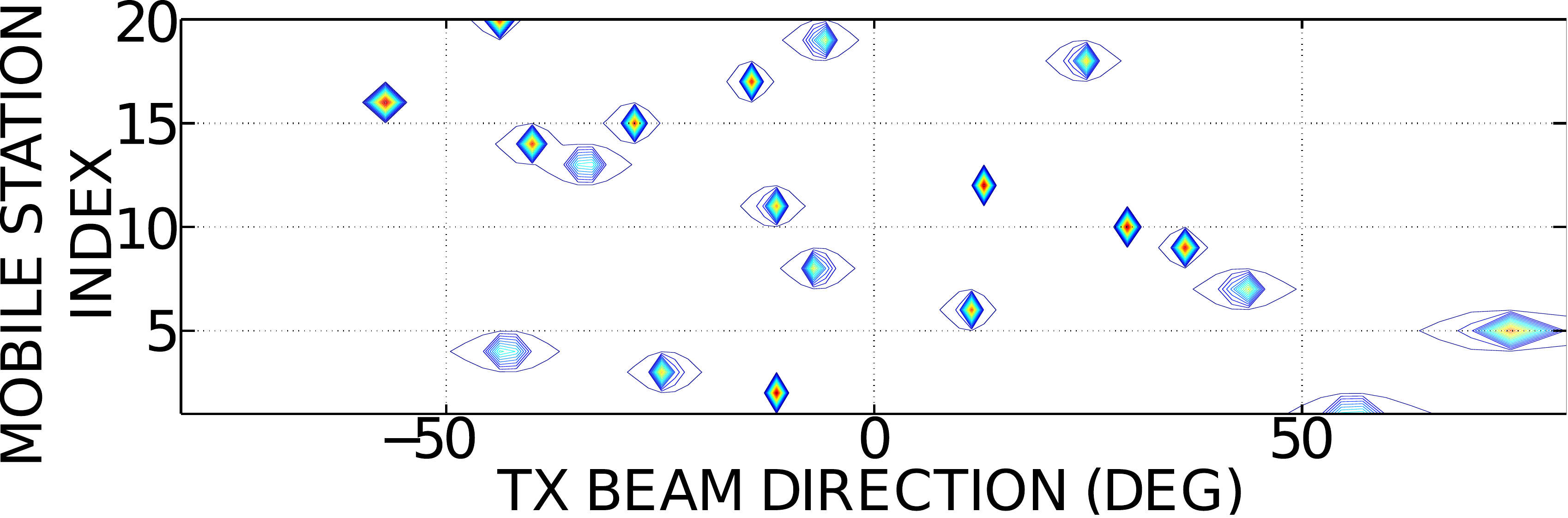}}
(a)\\
  \centerline{\includegraphics[width=2.5in]{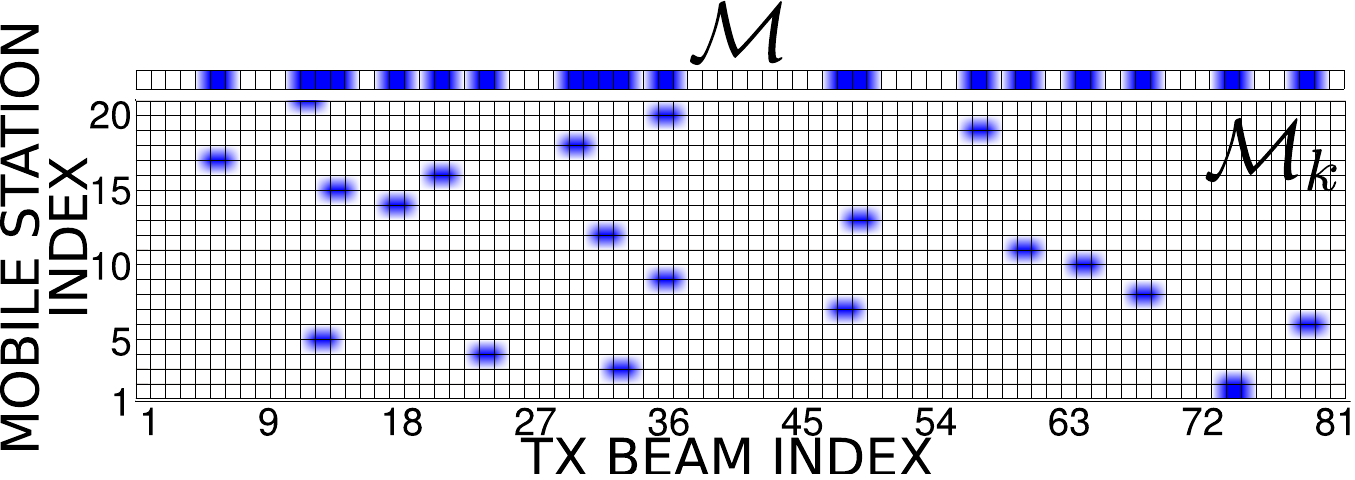}}
\\
(b)\\
\end{tabular}
\caption{(a) Contour plot of $|\bH_{\text{b}}^H|^2$ for a ULA with $N=81$, representing the {\em beamspace channel vectors} (rows)
for 20 MSs randomly distributed between $\pm 90^o$ (b) Illustration of beam masks $\cM_{k}$ and $\cM$ for the $\bH_{\text{b}}$ in (a).}
\label{fig:beam_sel}
\end{figure}

In \cite{brady:gcom13,brady:mu14}, an access point (AP) equipped with an $N$-dimensional ULA (or a lens-based front-end) that is communicating with $K$ single-antenna mobile stations (MSs) is considered. The multiuser channel is characterized by the $N \times K$ channel matrix $\bH$ where each column ($\bh_k$) corresponds to the channel vector for a different user.
The beamspace channel presentation is given by
\begin{equation}
\begin{split}
&\bH_{\text{b}} = \bU^*\bH = [ \bh_{b,1}, \bh_{b,2}, \cdots, \bh_{b,K} ] \ ; \ \\
&\bh_{b,k} = \bU^* \bh_{k} \ , \ k =1, 2, \cdots, K
\label{H_b_mu}
\end{split}
\end{equation}
where $\bh_{b,k}$ is the beamspace channel representation of the $k$-th MS.
An important property of $\bH_{\text{b}}$ is that it has a sparse structure representing the directins of the different MSs, as illustrated in Fig.~\ref{fig:beam_sel}(a). Each user, represented by $\bh_{b,k}$ is associated with a set of dominant beams as illustrated by rows in Fig.~\ref{fig:beam_sel}(a). These dominant beams define the beam masks $\cM_k$ for different users via a thresholding operation resulting in an overall beam mask $\cM$; see Fig.~\ref{fig:beam_sel}(b). The reduced complexity access point operates on these selected $p=|\cM| \ll N$ beams for precoding in the downlink and combining in the uplink.

The downlink system model is given by $\by = \bH^* \bx + \bv$
where $\by$ is the $K \times 1$ vector of received signals at the $K$ MSs, and $\bx$ is the $N \times 1$ is the transmitted signal. In a conventional (massive) MIMO system, a linear precoder takes the form $\bx = \bG \bs$, where $\bs$ is the vector of symbols for different MSs, and $\bG$ is the $N \times K$ precoding matrix that can be designed using various criteria, e.g. MMSE \cite{heath:07,nossek:05}. In beamspace, the downlink system model is given by $\by = \bH_{\text{b}}^* \bx_b + \bv \approx \tilde{\bH}_b^* \tilde{\bx}_b + \bv$, where $\bx = \bU_N \bx_b$, and the second equality represents the lower dimensional system characterized by $p \times K$ channel matrix $\tilde{\bH}_b$, and a corresponding $p \times K$ precoding matrix $\tilde{\bG}_b$; $\tilde{\bx}_b = \tilde{\bG}_b \bx$ \cite{brady:gcom13,brady:mu14}. The design of $\tilde{\bG}_b$ is computationally less intensive (compared to $\bG$) since $p \ll N$.

The uplink system model is given by $\by = \bH \bx + \bv$ where $\bx$ represents the vector of independent symbols from the $K$ MSs, and $\by$ represents the received signal at the access point. In a conventional MIMO system, the combiner operates on $\by$. In beamspace, the  combiner operates on $\by_b = \bU^*_N\by$, in particular on the $p$ dominant beams in $\tilde{\by}_b = \tilde{\bH}_b \bx + \tilde{\bv}$, thereby greatly reducing complexity as in the downlink case.

By capturing a sufficiently large fraction of channel power (via the choice of thresholds $\gamma_k$), the reduced-complexity linear beamspace precoders/combiners can be designed to deliver near-optimal performance \cite{brady:gcom13,brady:mu14}.
Using lens-based (or phase-shifter-based) front-end for analog beamforming can further reduce hardware complexity.
Integration of beam selection and multiuser channel estimation warrants further investigation.

\subsubsection{Multiuser precoding in the hybird precoding framework}

\begin{figure}[htb]
\centerline{\includegraphics[width=0.7\columnwidth]{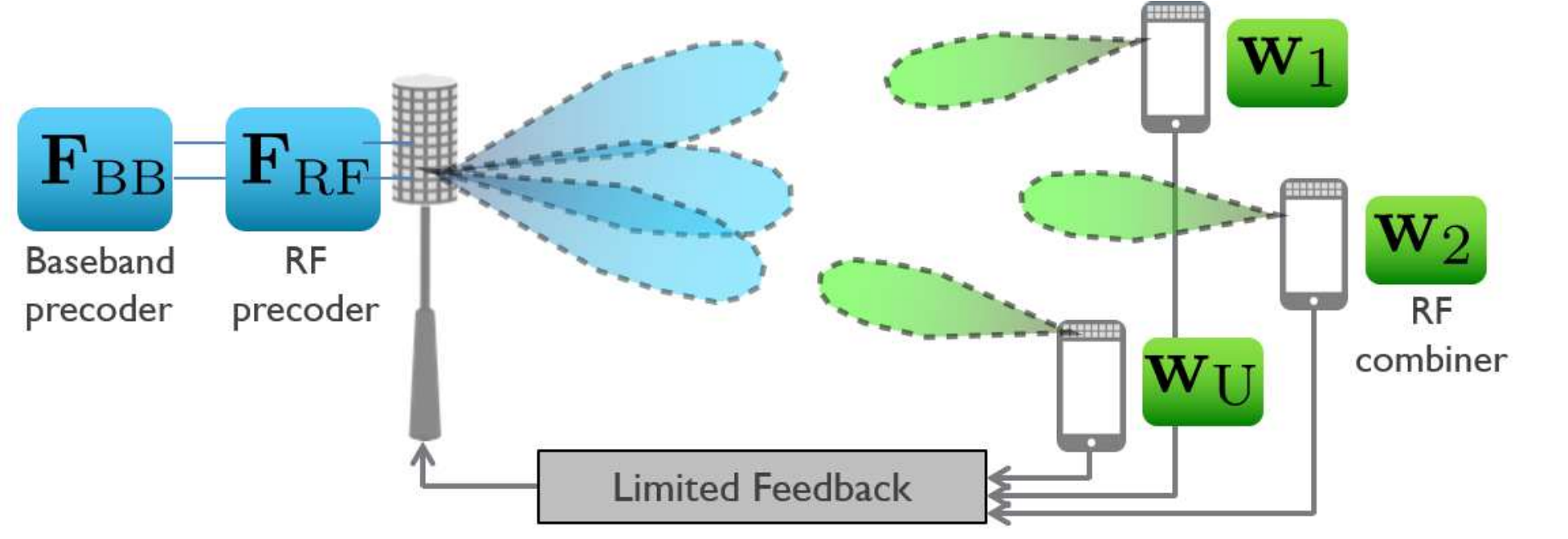}}
\caption{System model for the multiuser hybrid precoding design.}
\label{fig:MUhybrid}
\end{figure}

Hybrid precoding was also considered for multi-user mmWave systems \cite{Alkhateeb2014a}. In \cite{Alkhateeb2014a}, the downlink mmWave system was considered with the basestation employing hybrid analog/digital architecture and mobile users having analog-only combining (see Fig.~\ref{fig:MUhybrid}). For this system, a two-stage hybrid precoding algorithm was proposed and proved to achieve a near-optimal performance compared to a certain fully-digital approach. At the first stage, the analog beamformer and combiner are  designed to maximize the power at each user by single-user beam-training. At the second stage the baseband precoder is designed from the channel estimates performed at the users side to reduce inter-user interference. Only effective channels need to be trained, due to dimensionality reduction. 
The performance of multi-user mmWave systems with limited feedback, i.e. with quantizing both the analog and digital precoders, was also studied in \cite{Alkhateeb2014a}. It was shown that quantization of the baseband precoders is specially critical to preserve the hybrid precoding gain over analog-only beamsteering strategies. Further work is needed to develop hybrid precoding for both uplink and downlink with different precoding and combining strategies, and also for frequency selective channels. 

\section{Channel estimation at mmWave} \label{sec:train} \label{sec:compchan}
Channel estimates are useful for configuring the analog and digital beamformers that may be used in a mmWave system \cite{AMGH2014}. Conventional channel  MIMO channel estimation is difficult to apply in mmWave systems that use analog precoding and combining. The reason is that the channel measured in the digital baseband is intertwined with the choice of analog precoding and combining vectors  and thus the entries of the channel matrix can not be directly accessed. Further, a direct application of conventional channel estimation leads to the need to train many channel coefficients (due to the large number of elements in the transmit array) and long training sequences (due to the high bandwidth and low SNR prior to configuring the beamformer / combiners). This is problematic in applications where the channel varies quickly over time and must be estimated frequently. While beamtraining, as described in \secref{sec:beamtrain} can be used to avoid the need for explicit channel estimation, it does not necessarily provide enough information to implement more sophisticated transceiver algorithms, e.g. multiuser MIMO or interference cancellation, and it may take many iterations to find a good configuration. 

In  mmWave systems, leveraging channel sparsity is probably unavoidable. MmWave channels are sparse in both time and angular dimensions \cite{Wang2009,Sulyman2014}. Compressive adaptation techniques leverage mmWave channel spatial sparsity and overcome the limitations of codebook beamtraining. By using these techniques the estimation of the channel can be obtained from a small set of RF measurements. This section summarizes several approaches for channel estimation, leveraging sparsity to suggest reduced complexity implementations. The emphasis in this section is on single-shot estimators; adaptive estimators are an interesting topic for future work. 

\subsection{Sparse channel estimation for hybrid architectures based on phase shifters or switches}\label{sec:sparsehybrid}

Channel estimation at mmWave can be formulated as a sparse problem  where  the measurement matrices are the hybrid precoders/combiners. Different ideas can be  employed to design these measurement matrices: (i) ideas based on adaptive compressed sensing \cite{Malloy2012,Malloy2012a,Iwen2012, Alkhateeb2014}; and (ii) ideas that rely on traditional random compressed sensing using pseudorandom weights in a phased array  \cite{Ramasamy2012a,Ramasamy2012b,Berraki2014,Alkhateeb2015} or pseudorandom antenna switching \cite{Roi2015}. 

First we explain the general approach described in \cite{Alkhateeb2014} based on the hybrid architecture in Fig.~\ref{fig:arch-hybrid}. Suppose that $\Mr$ measurements at successive $\Mr$ instants using $\Mt$ different beamforming vectors at $\Mt$ time slots are performed. Let $\bX$ be a diagonal matrix containing the $\Mt$ training symbols on its diagonal,  $\bF_{\text{t}}=[\bff_1,\bff_2,\ldots,\bff_{\Mt}]$ be the training precoding matrix  of size $\Nt \times \Mt$,   $\bW_{\text{t}}= [\bw_1,\bw_2,\ldots,\bw_{\Mr}]$ be the $\Nr \times \Mr$  training combining matrix, and denote  $\bQ$ as the $\Mr\times \Mt$  noise matrix. Since we are considering a hybrid architecture $\bF_{\text{t}}=\bF^{\text{t}}_\text{RF}\bF^{\text{t}}_\text{BB}$ and  $\bW_{\text{t}}=\bW_\text{RF}^{\text{t}}\bW_\text{BB}^{\text{t}}$. The matrices correponding to the analog configuration $\bF_\text{RF}^{\text{t}} \in \mathbb{C}^{\Nt\times \Mt}$, and   $\bW_\text{RF}^{\text{t}} \in \mathbb{C}^{\Nr\times \Mr}$ are assumed to have constant modulus entries and represent the RF precoding/combining matrices while $\bF_\text{BB}^{\text{t}} \in \mathbb{C}^{\Mt\times \Mt}$
and $\bW_\text{BB}^{\text{t}} \in \mathbb{C}^{\Mr\times \Mr}$ represent the baseband  precoding/combing matrices with a block diagonal structure. Concatenating the $\Mr$ received vectors,  he $\Mr \times \Mt$ received signal can be written as
\begin{equation}
\bY=\bW_{\text{t}}^*\bH\bF_{\text{t}} \bX+\bQ,
\label{eq:receivedsignalmodel}
\end{equation}
Note that the notation is different than in \secref{sec:hybridprecode}. In the channel estimation case the preferred beamforming directions are not yet available; therefore multiple measurements are needed over time, requiring the use of different beamforming and combining matrices. 

Assuming that all transmitted symbols are equal, and using the extended virtual channel model in (\ref{eq:extendedvirtual}) with quantized AoAs/AoDs, the received signal after vectorization can be approximated by \cite{Alkhateeb2014}
\begin{equation}
\by_v=\sqrt{P}(\bF_{\text{t}}^T \otimes \bW_{\text{t}}^*)\bA_D\bh_\text{b}+\bn_{\bQ},
\label{vecy}
\end{equation}
where $\bh_\text{b}=\text{vec}(\tilde{\bH}_\text{b})$ is a $G^2 \times 1$ sparse vector which contains the path gains of the quantized spatial frequencies. Each column  of the $\Nt\Nr\times G^2$ dictionary matrix 
$\bA_D$ represents the Kronecker product  $\mathbf{a}_{\text{T}}^*(\phi_k) \otimes \mathbf{a}_{\text{R}}(\theta_j)$, where $\phi_k$ and $\theta_j$ are the $k$-th and $j$-th points of the uniformly quantized grid of $G$ points, with $G\gg N_p$.  Using Kronecker product properties 
an alternative expression is
\begin{equation}
\by_v=\sqrt{P}(\bF_\text{t}^T  \bar{\bA}_{\text{T}}^*\otimes \bW_\text{t}^*\bar{\bA}_{\text{R}})\bh_\text{b}+\bn_{\bQ}.
\label{vecy2}
\end{equation}

The channel estimation problem is formulated as a non-convex combinatorial problem
assuming that $\mathbf{h}_\text{b}$ is a sparse vector,
\begin{equation}
\label{SMV}
	\underset{\mathbf{h}_\text{b}}{\min} \  \| \mathbf{h}_\text{b} \|_0 \;\;\; \text{subject to} \;\;\; \| \mathbf{y}_v-\sqrt{P}(\mathbf{F}_{\text{t}}^T \otimes \mathbf{W}_{\text{t}}^*)\mathbf{A}_D\mathbf{h}_\text{b}\|_ 2 \leq \sigma .
\end{equation}
Given this sparse problem, compressed sensing tools can be employed to solve it. In \cite{Alkhateeb2014}, an adaptive compressed sensing based solution was proposed to iteratively estimate the mmWave channel paths. Alternatively, standard greedy recovery algorithms, such as Orthogonal Matching Pursuit (OMP), can be used to solve \eqref{SMV} efficiently. 
The matrix $(\mathbf{F}_{\text{t}}^T \otimes \mathbf{W}_{\text{t}}^*)\mathbf{A}_D$ plays a key role in establishing recovery guarantees.
 Note that $\bA_D$ functions as the sparsifying dictionary and $(\bF_{\text{t}}^T \otimes \bW_{\text{t}}^*)$ works as a measurement matrix that needs to be efficiently designed using compressed sensing theory to guarantee the success of the sparse reconstruction. The aim is to design training sequences of precoding/combining vectors that define a sensing matrix providing low coherence. Next, we explain in more detail these compressive approaches when using different analog processing hardware.
One limitation of compressive channel estimation strategies at the receiver is the algorithms usually assume knowledge of the array geometry employed at the transmitter side, which may not be available in practice.


In \cite{Lee2014} a hybrid architecture based on phase shifters and the received signal model in (\ref{vecy}) is also assumed. The sparse recovery problem  in (\ref{SMV}) is solved for a given sparisty of the channel using a multigrid OMP approach. The algorithm starts with a coarse grid which is iteratively refined only around the regions corresponding to the coarse AoAs and AoDs.  From (\ref{vecy2}), we can define the {\em equivalent} measurement matrix as $\boldsymbol{\Phi}=\sqrt{P}(\bF_{\text{t}}^T \bar{\bA}_{\text{T}}^*\otimes \bW_{\text{t}}^*\bar{\bA}_{\text{R}})$. Since the grid is iteratively defined, $\bar{\bA}_{\text{T}}$ and $\bar{\bA}_{\text{R}}$, which work as the dictionary matrices, are  different at each step of the reconstruction algorithm.  
The RF beamforming/combining  training vectors $\bF_\text{RF}^{\text{t}}$ and $\bW_\text{RF}^{\text{t}}$ are chosen as the columns of the  $\Mt\times \Mt$ and $\Mr\times \Mr$ DFT matrices. The baseband  precoding/combining training vectors $\bF_\text{BB}^{\text{t}}$ and $\bW_\text{BB}^{\text{t}}$ are designed to minimize the coherence of the initial equivalent measurement matrix.

A hybrid architecture based on phase shifters constrains the RF precoding/combining matrices to have unit norm entries.
An architecture based on switches restricts each column of $\bF^{\text{t}}_\text{RF}$ and $\bW^{\text{t}}_\text{RF}$ to have exactly a one at the index of the selected antenna and zeros elsewhere.
In \cite{Roi2015}, it was shown that analog-only  binary pseudorandom combining matrices based on switches provide equal or even lower coherence than measurement matrices associated to an architecture based on phase shifters. Besides of having a similar channel estimation performance, hybrid architectures based on switches lead to a lower  power consumption  with respect to phase shifters.

The contributions summarized  above show the success of compressive channel estimation in simple mmWave systems. Many open problems remain.  To further increase the performance of sparse recovery algorithms, it would be interesting to design  alternative training precoders/combiners at RF and baseband that minimize the coherence of the equivalent measurement matrix. It is also interesting to analyze the trade-offs between the training length and the number of RF chains for the different architectures. The design of limited feedback strategies for the mmWave MIMO channel is also interesting, as the estimators and quantizers are intertwined. Estimating the array geometry at the same time as the channel is another challenging direction, as is feedback and feedforward of array geometry information. Finally,  it would be interesting to formulate the channel estimation problem for a multi-cell system and a wideband channel model, to study the influence of the inter-cell interference into the performance of compressive channel estimators.

\subsection{Beam training and sparse channel estimation in lens-based CAP-MIMO transceivers}
\label{sec:est_capmimo}
Consider and $\Nr \times \Nt$ mmWave MIMO system with a lens-based transceiver architecture such as CAP-MIMO. Channel estimation consists of two steps: i) determining the {\em channel beam masks}, $\cM$, $\cM_{\text{t}}$ and $\cM_{\text{r}}$, defined in Sec.~\ref{sec:beam_masks}, that determine the low-dimensional beamspace channel matrix ${\tilde \bH}_{\text{b}}$, and ii) estimation of the entries of ${\tilde \bH}_{\text{b}}$. The second step can be accomplished by sequentially exciting the transmit beams in $\cM_{\text{t}}$ and, for each excited transmit beam, measuring the corresponding receive beams in $\cM_{\text{r}}$. This yields a columnwise estimate  ${\tilde \bH}_{\text{b}}$ \cite{kotecha:04}. The determination of $\cM$ essentially boils down to sequential transmission and thresholding: sequentially exciting different transmit beams, and determining the receive beams with sufficiently high power for each transmitted beam. This approach generally requires somewhere between $\cO(N)$ and $\cO(N^2)$ transmissions, depending on the number of simultaneous measurements possible at the receiver. While many different algorithms can be developed, the choice of the threshold in determining dominant channel entries is key.

\subsection{Channel estimation with 1-bit architectures} \label{sec:onebitchanest}

Channel estimation with one-bit ADCs for the MIMO channel in general \cite{Mezghani_WSA10,Dabeer_ICC10} and in the context of mmWave \cite{OneBitMo2014} is surprisingly effective when understood from a mathematical perspective. In \cite{OneBitMo2014}, channel sparsity is exploited and the narrowband virtual channel model in (\ref{eq:extendedvirtual}) is considered, which allows for a sparse recovery problem to be formulated.  The received signal using this particular architecture can be written as
\begin{equation}
\bY=\text{sign}(\bH \bF_{\text{t}} \bX+\bQ),
\end{equation}
where $\bX$ is the training sequence and $\bQ$ is the  i.i.d. Gaussian noise. Using the virtual channel representation in (\ref{sys_bs}), setting $\bF_{\text{t}}=\bU_{\Nt}$, using the training sequence $\bX=\bF_{\text{t}}\bZ$
\begin{equation}
\text{vec}(\bY)=\text{sign}((\bZ^T\otimes\bU_{\Nr})\text{vec}(\bH_{\text{b}})+\text{vec}(\bQ)).
\label{receivedsignalmodel-1bit}
\end{equation}
The problem of estimating  $\bh_{\text{b}}=\text{vec}(\bH_{\text{b}})$ given $\bZ$, $\bU_{\Nr}$ and the received signal can be solved using the one-bit compressive sensing framework introduced in \cite{Jacques_IT13} to recover sparse vectors. The reconstruction can be further improved if prior information about the distribution of $\bh_{\text{b}}$ is used \cite{OneBitMo2014}. In this case, the generalized approximate message passing (GAMP) algorithm can be used to solve the optimization problem in a small number of steps.
 
  
Channel estimation of the broadband channel is an active area of research. The closed-form ML estimator of the channel can be derived for the one-tap SISO channel \cite{Mezghani_WSA10}, but it is intractable for frequency-selective channels. Prior work proposed to transmit periodic bursty training sequences and estimate each tap of the channel responses separately\cite{Dabeer_ICC10, Zeitler_TSP12}. A more efficient way is to include the correlation of the channel responses (for instance, the sparsity of the mmWave channel \cite{OneBitMo2014}). The GAMP algorithm is also appealing in this case \cite{Mezghani_WSA12,OneBitMo2014,Wen_ChaoKaoArxiv15}.

 \subsection{Multiuser channel estimation}
 In \cite{Alkhateeb2015}, a compressed-sensing based multi-user mmWave system operation was proposed in which the basestation and mobile users employ random beamforming/measurement matrices to estimate the downlink channel parameters (AoAs/AoDs and path gains). Then, quantized AoA/AoD knowledge is fed back to the basestation, which uses this to construct the data transmission beamforming vectors. Apart from adaptive compressed sensing, random compressed sensing may be more suitable for multi-user systems as all the mobile users can simultaneously estimate their channels thanks to the randomness nature of the transmitted beams. One important question when random compressed sensing tools are used to estimate mmWave channels is how many measurements are need? To give an initial answer to this question, \cite{Alkhateeb2015} derived a simple expression for the per-user achievable rate as a function of the number of compressed sensing measurements in some special cases. It was shown that  at least an order of magnitude fewer compressed sensing measurements are needed compared with exhaustive search solutions. Further work is  needed to develop multi-user channel estimation strategies for hybrid precoding, low resolution ADCs, and broadband channels. 



\section{Conclusions}\label{sec:conclusion}

%
Communicating at mmWave is not simply a matter of just changing the carrier frequency. Going to mmWave changes the assumptions that underly prior developments in signal processing for communication. The radio frequency hardware introduces constraints that have ramifications on the beamforming, precoding, and channel estimation algorithms. The propagation channel has higher dimension, with more spatial sparsity, different pathloss characteristics, and extreme sensitivity to blockage. Large antenna arrays may be used for both transmission and reception, renewing the importance of MIMO communication.  There are many open research problems relating to channel modeling, precoding, receiver design, channel estimation, and broadband channels, not to mention system design challenges that arise when mmWave is used in personal area networks, local area networks, cellular networks, vehicular networks, or wearable networks. There is a bright future ahead in signal processing for mmWave wireless systems.


\end{document}